\RequirePackage{fix-cm}
\documentclass[smallextended]{svjour3}       
\smartqed  
\usepackage{epsfig}
\usepackage{epstopdf}
\usepackage[T1]{fontenc}
\usepackage{amssymb}
\usepackage{amsmath}
\usepackage{amsfonts}
\usepackage{verbatim}
\usepackage{graphicx}
\usepackage{hyperref}
\usepackage[usenames,dvipsnames]{xcolor}
\usepackage{algorithmic}[5]
\usepackage{algorithm}
\usepackage{array}
\usepackage{multirow}
\usepackage{caption}
\usepackage{subcaption}
\usepackage{multicol}
\usepackage{pgfplots}
\usepackage{caption}
\usepackage{subcaption}
\usepackage{slashbox}
\usepackage{tikz-timing}[2009/05/15]
\DeclareCaptionType{timingdiag}[Timing diagram][List of Timing Diagrams]
\begin{document}

\title{Accelerating More Secure RC4 : Implementation of Seven FPGA Designs in Stages upto 8 byte per clock}

  \author{Rourab Paul,
        Hemanta Dey,
        Amlan~Chakrabarti,
        and~Ranjan~Ghosh}


\institute{R. Paul, Amlan Chakrabarti, Ranjan Ghosh \at
             IT Dept, University of Calcutta, India \\
                    \email{rourabpul@gmail.com}           
           \and
          H. Dey \at
            Techno India University
}

\date{Received: date / Accepted: date}

\maketitle

\begin{abstract}

RC4 can be made more secured if an additional RC4-like Post-KSA Random Shuffling (PKRS) process is introduced between KSA and PRGA.  It can also be made significantly faster if RC4 bytes are processed in a FPGA embedded system using multiple coprocessors functioning in parallel. The PKRS process is tuned to form as many S-boxes as required by particular design architectures involving multiple coprocessors, each one undertaking byte-by-byte processing. Following a recent idea \cite{springerlink:one_byte} \cite{ieee:two_byte} the speed of execution of each processor is also enhanced by another fold if the byte-by-byte processing is replaced by a scheme of processing two consecutive bytes together. Adopting some new innovative concepts, three hardware design architectures are proposed in a suitable FPGA embedded system involving 1, 2 and 4 coprocessors functioning in parallel and a study is made on accelerating RC4 by processing bytes in byte-by-byte mode achieving throughputs from 1-byte-in-1-clock to 4-bytes-in-1-clock. The hardware designs are appropriately upgraded to accelerate RC4 further by processing 2 consecutive RC4 bytes together and it has been possible to achieve a maximum throughput of 8-bytes per clock in Xilinx Virtex-5 LX110t FPGA \cite{xilinx:online}  architecture followed by secured data communication between two FPGA boards.
\end{abstract}

\section{Introduction}
The RC4 became widely popular over the last three decades because of its simple and straight forward algorithm. The RC4 was proposed by Ronald Rivest in 1987 \cite{rc4:source} as the first secret commercial stream cipher for rendering security services by RSA Data Security \cite{rc4:main} \cite{ronald}. In 1994 an anonymous insider \cite{mail} made the RC4 algorithm public following which it attracted attentions of many researchers. Presently, RC4 is a part of many network protocols, e.g. SSL, TLS, WEP, WPA and many others. There were many cryptanalysis to look into its key weaknesses  \cite{roos1}, \cite{alex} \cite{fluhrer} \cite{DBLP:spaul} \cite{springerlink:gpaul} followed by many new stream ciphers \cite{t:Good}\cite{p:leg} \cite{p:kitsos} \cite{m:gal}, \cite{port}. RC4 is still the popular stream cipher since it is executed fast and provides reasonably high security  \cite{shuvomoy:book}. 
\par The underlying concept of RC4 algorithm has been derived from the Knuth's idea \cite{knuth} of shuffling of two numbers among some finite set of numbers; one number is sequentially chosen and the other one is chosen randomly based on a uniformly distributed random number generator between zero and unity. Ronald Rivest chose the complete set of 8-bit numbers as the finite set of numbers and stored them in an identity S-box. For the purpose of adopting Knuth's idea of shuffling, he innovatively adopted a simple method of modular addition to choose a random element which is shuffled with another element chosen sequentially. The algorithm has two stages of operation, the first stage, named as KSA (Key Scheduling Algorithm) is undertaken 256 times in each of which a key element always plays a role in the modular addition that chooses the random element and in the second stage, named as PRGA (Pseudo Random Generation Algorithm) and undertaken infinitely, the random element used for shuffling is chosen without any key element in the modular addition and random stream bytes are generated. It was Roos \cite{roos1} in 1995 who first observed key weakness in RC4 by marking some weak keys and following a detailed analysis noted that the algorithmic simplicity provides clues to predict few initial key bytes based on few initial stream bytes, supposed to be random. As soon as PRGA starts Roos~\cite{roos1} intended that the PRGA process continues with its swapping only for some finite number of times with an expectation that the arrangement of S-box elements would be better random which would be able to give sequence of stream bytes without any key bias and thereby suggested to discard about 256 initial PRGA bytes. Paul and Preneel \cite{DBLP:spaul} who after marking some other weak keys innovatively adopted RC4 algorithm for two identity S-boxes using two different keys, two consecutive stream bytes are generated simultaneously from two S-boxes in one loop breaking the sequence the stream bytes supposed to have in conventional RC4, thereby expecting to eliminate the key bias observed by Roos \cite{roos1} and also making RC4 faster. However, on observing weakness in statistical randomness in stream bytes, Paul and Preneel suggested to undertake 256 times of discarding PRGA bytes which amounts to discard 512 PRGA bytes. There were studies on statistical weakness in RC4 stream bytes\cite{ronald}, related-key cryptanalysis of RC4 \cite{roos1} and weakness in key scheduling algorithm of RC4 \cite{alex}.  Adopting a method to calculate probabilities of post-KSA PRGA stream bytes, Paul and Maitra \cite{shuvomoy:book} have analyzed RC4 in greater detail, could mark those stream bytes showing probabilities lower than the desired one and could relate them with appropriate key characters.  In order that probabilities of each of the entire 8-bit 256 characters that PRGA stream bytes generate becomes equal to 1/256, they proposed to add two layers of permutations in KSA itself so that the arrangement of S-box elements becomes better random making the algorithm little more complex.  Later on, in their book \cite{springerlink:gpaul} they suggested discarding first 1024 PRGA bytes instead of strictly resorting to more complex algorithm. It is visualized that there is a serious advantage if the activity of discarding PRGA bytes is made formal and a Post-KSA Random Shuffling (PKRS) process is proposed between KSA and PRGA in which no key elements are considered and which is being run 1024 times in order to form as many S-boxes as would be required by a particular design architectures involving many coprocessors in FPGA in order to make RC4 more secured as well as faster.  

\par The first RC4 hardware design was reported in 2003 \cite{patent:matthews} where 1-byte was processed in 3-clocks and the same was implemented in ASIC. Another design with identical throughput was also implemented in FPGA in 2004  \cite{IEEE:b}. In 2008, a hardware design using a pipelining architecture was proposed where 2-bytes are sequentially processed within 2-clocks and the same was implemented in FPGA \cite{dp:math}. Till date the fastest known hardware design of RC4 is the processing of 2-bytes together in 1-clock reported in 2010 \cite{springerlink:one_byte}, whose ASIC implementation was made in 2013 [2]. A 3-stage design scheme following a suitable pipelining architecture was proposed in \cite{springerlink:one_byte} for hardware processing of 2-bytes in 2-clocks and a 2-stage design scheme following a suitable pipelining architecture was proposed in \cite{ieee:two_byte} for hardware processing of 2-bytes in 1-clock. 
\par The underlying idea of the present paper is to optimally use parallelization feature of FPGA, to introduce innovative design concepts in it, to systematically study simulation results of few design ideas by effectively using available hardware resources in order to achieve maximum possible throughput. Following this motivation, seven hardware designs, from 1-byte in 1-clock to 8-bytes in 1-clock are implemented in FPGA and a comparative study is made concerning throughput, power consumption, resource usage and statistical randomness of random key streams. At the backdrop of all the seven designs, there are five important design concepts among which the first one is the novel concept of processing 2-bytes together as proposed in \cite{springerlink:one_byte} with a little modification and the rest four are the new ones. Briefly the five concepts are, (i) 2-bytes are processed together in FPGA in 1-clock in which the 3-stage pipelining architecture proposed in \cite{springerlink:one_byte} is replaced by a 2-stage one, (ii) multiple coprocessors are used, each one coupled with an 8-bit S-box made capable to process 1-byte in 1-clock or 2-bytes together in 1-clock, (iii) swapping is executed by using a MUX-DEMUX combination replacing the use of temporary variable, (iv) data are processed during both the rising and falling edges of a clock instead of using one of its two edges, (v) the KSA and PGRA processes of RC4 are implemented together in the same silicon slice as one composite block instead of implementing them in two different silicon areas. The first four issues are responsible to increase throughput, while the fifth one is responsible to reduce both power consumption and resource usage. All the five concepts are appropriately incorporated in all the designs as required and the encryption and decryption engines of each of all the modified RC4 designs are embedded respectively in two FPGA boards in stages as coprocessors and the communication between them has been achieved for all these designs using Ethernet. The study indicates that one can implement various hardware designs in advanced FPGA systems right in the laboratory and can achieve throughput of Gbps order which happens to be possible only in ASIC technology in recent past. It seems that many applications including crypto processors implemented in the FPGA are  turning out to be competitive with ASIC not only from the cost point of view but also from the technology angle.   
\par The introduction is little elaborated by briefly describing the RC4 algorithm in Sec. 1.1 and the organization
of presentation of the paper in Sec.1.2.

\begin{figure}[!htb]
\centering
\vspace{-10pt}
\includegraphics[scale=0.31]{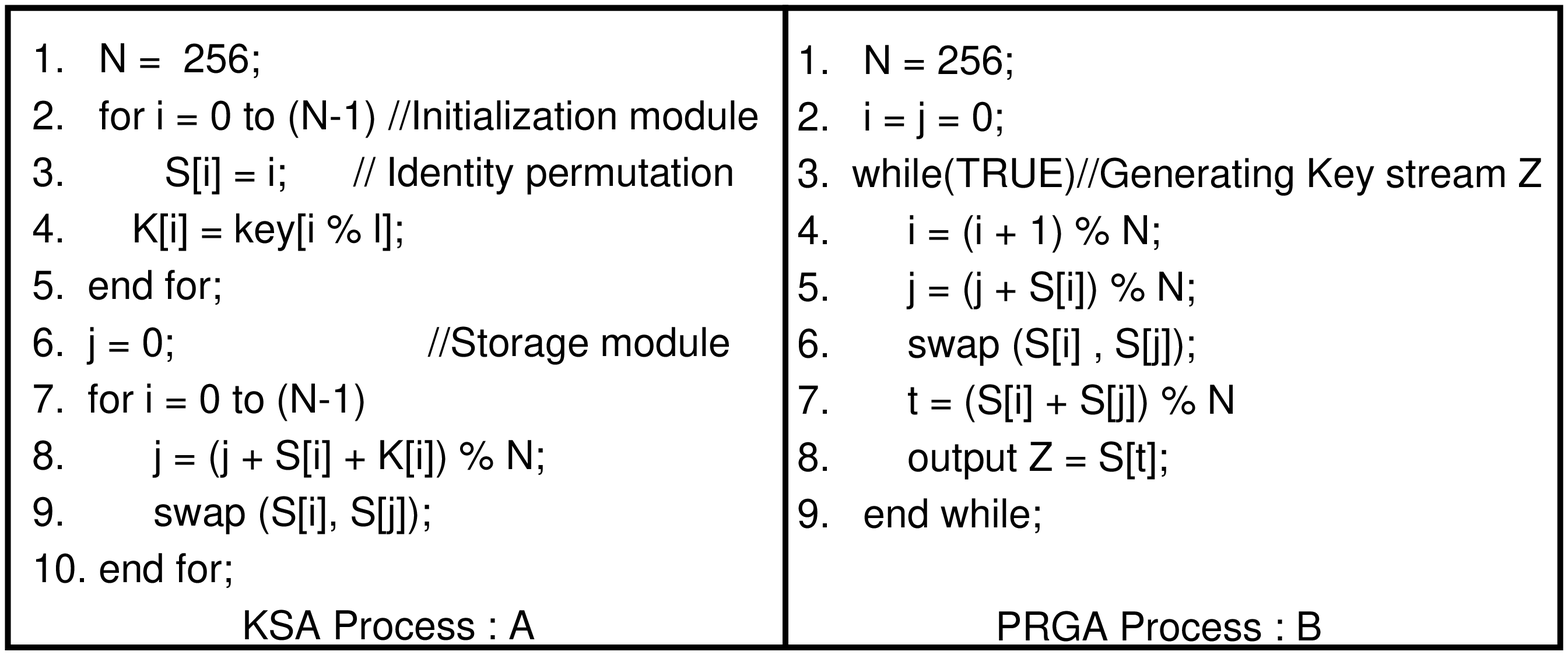}
\vspace{-10pt}
\caption{RC4 Algorithm}
\label{fig:ksa_prga_fig}
\vspace{-40pt}
\end{figure}
\subsection{RC4 Algorithm in Brief}
The RC4 algorithm is briefly described in Fig.\ref{fig:ksa_prga_fig} mentioning separately a KSA (Key Scheduling Algorithm) process in Fig.1(A) and a PRGA (Pseudo Random Generator Algorithm) process in Fig.1(B). The novelty and simplicity of RC4 lies both in KSA and PRGA.  The KSA consists of two loops, (1) the first one is the initialization loop executed between lines 2 and 5 performing two activities  (a) formation of an 8-bit identity S-box (line 3) and (b) storing the given key characters repetitively in an 8-bit K-box (line 4) and (2) the second is the randomization loop executed between lines 6 and 10 involving two indices $i$ and $j$ performing three activities as, (a) both the indices start from zero (lines 6 and 7), (b) $i$ is sequentially increased (line 7) while $j$ is randomly upgraded  by modulo 256 addition of $j$ itself with the data elements fetched from the S-box and the K-box corresponding to $i$ (line 8) and (c) the identity S-box is randomized by continuously shuffling its two elements corresponding to the indices $i$ and $j$ (line 9). The randomization loop in KSA is continued till the processing of the last key character in the K-box.  Once KSA ends, the PRGA makes the two indices $i$ and j to start from zero and executes an infinite loop shown between lines 3 and 9 of Fig. 1(B).  There are five issues in the infinite PRGA loop, (a) $i$ starts from i=1, not from i=0 which is accessed to fetch its data only when i=255 (line 4), (b) the upgradation of random index $j$ follows the KSA randomization process without K-box elements (line 5), (c) the randomization of S-box is continued by continuous shuffling as in KSA (line 6), (d) a temporary random index $t$ is computed by modulo 256 addition of two S-box elements corresponding to $i$ and $j$ and (e) the S-box data element corresponding to the temporary random index $t$ contributes to the stream of random key bytes PRGA produces. The idea of KSA is to use the given key characters in randomizing the initial identity S-box elements and to present a given-key-dependent randomized S-box to the PRGA. The idea of PRGA is to continue randomizing the S-box elements without the given key and to contribute one of its elements as a PRGA-byte to the stream of random key bytes.
\vspace{-15pt}
\subsection{Organization of the paper}
In Sec. \ref{sec2} the various hardware issues involved in RC4 implementation are reviewed including  the technique adopted to process two consecutive bytes together.  The conceptual milestones conceived in connection to our designs have been well described in Sec. 3 along with implementations of seven designs. The results of different implementations regarding consumptions of hardware resources and electrical power including a comparative study of the throughput of all the seven designs with the existing designs are described in Sec. \ref{ram}. The results of statistical tests are presented in Sec.\ref{nist}. The paper is concluded in Sec. \ref{con}.
\vspace{-20pt}
\section{Brief review of Hardware implementations of RC4 and other advancements}
\vspace{-10pt}
\label{sec2}
The ASIC and FPGA implementations of RC4 are reviewed in Sec.2.1. The Novel idea of processing  two consecutive PRGA bytes together is presented in Sec. 2.2.  Various methods adopted in swapping of  two elements are discussed in Sec. 2.3.  The review of timing analyses of few designs that  have already  been implemented are being made in Sec. 2.4.
\vspace{-20pt}
\subsection{Review of existing Hardware implementations of RC4}
\label{exist}
In 2004  Kitsos et. al. \cite{IEEE:b} implemented hardware architecture to process 1 RC4 byte in 3 PRGA clocks. During 1st clock the architecture computes sequential index $i$ and the random index $j$, during 2nd clock it retrieves $S[i]$ and $S[j]$ from RAM, adds them and stores them in register $t$ and during the 3rd clock, it swaps $S[i]$ and $S[j]$, accesses $t$ and retrieves $S[t]$ as $Z$. It may be noted that as the swapping and retrieving $S[t]$ as $Z$ are executed during the same 3rd clock, there remains a scope for breakdown of the design algorithm at an instant when the value of $t$ turns out to be either of $i$ or $j$. In 2003 Matthews \cite{patent:matthews}, proposed another 3-byte in 1-clock architecture where KSA and PRGA both consume 3 clocks for each iteration. In 2008 Matthews again proposed a new 1-byte per clock pipelined architecture \cite{dp:math}. Lastly, S. Sen Gupta et. al. proposed two architectures in \cite{ieee:two_byte} and \cite{springerlink:one_byte}. In \cite{ieee:two_byte}, a loop unrolled architecture is proposed to process 2 bytes in 2 consecutive clocks. In \cite{ieee:two_byte}, following a pipelined architecture, hardware implementation of the design proposed in \cite{springerlink:one_byte} is presented.
\vspace{-20pt}
\subsection{Novel technique to process two consecutive PRGA bytes of RC4 together}
\label{novel:tech}
In an $n^{th}$ loop of Fig. \ref{fig:ksa_prga_fig}~B, $i_n$ is calculated from $i_{(n-1)}$ using line 4, using line 5 $j_n$ is calculted from $j_{n-1}$ and $i_n$ and then $Z_n$ is computed using lines 7 and 8. In the $(n+1)^{th}$ loop, one is supposed to compute $i_{n+1}$ and $j_{n+1}$ followed by $Z_{n+1}$ from $j_n$ and $i_n$. If it is intended to compute $Z_n$ and $Z_{n+1}$ together in the $n^{th}$ loop, it is necessary to calculate $i_n$, $j_n$, $i_{n+1}$ and $j_{n+1}$ from $i_{n-1}$ and $j_{n-1}$ using Eqs. \ref{in_equ} and \ref{jn_equ} for all possible eventualities of their equalities and/or inequalities. 
\vspace{-5pt}
\begin{equation}
\label{in_equ}
i_n=i_{n-1}+1,
\vspace{-10pt}
i_{n+1}=i_n+1=i_{n-1}+2
\end{equation}
\vspace{-10pt}
\begin{equation}
\label{jn_equ}
j_n=j_{n-1}+S_{n-1}[i_n]
\end{equation}
Before the computation of $i_{n+1}$ and $j_{n+1}$, the data of $S_{n-1}[i_n]$ and $S_{n-1}[j_n]$ is expected to be mutually swapped following which the $S_{n-1}$ is renamed as $S_n$.  Even though swap does not take place, its effect should be considered while computing $i_{n+1}$ and $j_{n+1}$.  If $i_{n+1}\neq j_n$, the swap has no effect on the computation of $j_{n+1}$ and one can consider  $S_{n-1}[i_{n+1}]$ as $S_n[i_{n+1}]$. If $i_{n+1}~=~j_n$, the swap influences the computation of $j_{n+1}$ and one should consider  $S_{n-1}[i_{n}]$ instead of $S{n}[i_{n+1}]$, since the data located at $S_{n-1}[j_n]$ is supposed to be at $S_{n-1}[i_n]$. Considering the conditions of inequality and equality between $i_n$ and $j_n$, the eq. (2) is re-written as,   
\vspace{-5pt}            
\begin{align}
\label{jn+1_equ}
j_{n+1}& = j_n + S_n[i_{n+1}] = j_{n-1} + S_{n-1}[i_n] + S_n[i_{n+1}] \nonumber\\
&= j_{n-1} + S_{n-1}[i_n] + S_{n-1}[i_{n+1}], ~~~if ~i_{n+1}\neq j_n \nonumber\\
&= j_{n-1} + S_{n-1}[i_n] + S_{n-1}[i_n], ~~~if ~i_{n+1}=j_n
\end{align}
\vspace{-2pt}
Of the four variables, $i_n,~ i_{n+1},~ j_n,~and~ j_{n+1}$, there would be 6 pairs out of which it is not necessary to consider the pair $(i_n, i_{n+1})$ since the pair is always unequal.  Hence, two sets each of 5 conditional pairs are considered below as follows, \\
(A) 5 conditions of inequality:\\
$i_{n+1}\neq j_{n+1},~~~~~~~ i_{n+1}\neq j_n, ~~~~~~~ i_n\neq j_{n+1},$\\
7(B) 5 conditions of equality :\\
$i_{n+1}= j_{n+1},~~~~~~ i_{n+1}=j_n,~~~~~~ i_n=j_{n+1},$\\
$i_n = j_n ~~~and~~~ j_{n+1}= j_n$ \\
In (B) the two conditions together ($ i_{n+1}$ = $ j_n$) and ($i_n$ = $j_n$) and another two conditions together ($i_{n+1}$ =$  j_{n+1}$) and ($i_n$ = $ j_{n+1}$) are redundant since $ i_{n+1} \neq i_n$.  It is sufficient to consider one of the two in each, i.e. ($i_{n+1} = j_n$) and ($i_n$ = $j_{n+1}$) in (B) as well as in (A)  with inequality.  Hence  there will be two sets each with 3 effective conditional pairs 
as stated below, \\
(A)  3 conditions of inequality:\\
$i_{n+1} \neq j_n, ~~~~~i_n\neq j_{n+1} ~~~~and ~~~~~j_{n+1}\neq j_n$\\
(B)  3 conditions of equality :\\
$i_{n+1}=j_n, ~~~~~i_n=j_{n+1} ~~~~~and~~~~~ j_{n+1}= j_n$\\
Out of the two sets each with 3 logical conditions of inequalities and equalities, there would be eight possible combinations of logical conditions which are presented in Table \ref{z2_table} along with data movement during 2 swaps and computations of $Z_n$ before the swap and $Z_{n+1}$ after the swap. 8 different conditions of data movement are shown in Table \ref{z2_table}. In the present computation $i_n$ cannot be equal to $i_{n+1}$, since that leads to an impossible situation of $i_n = i_{n+1}$ shown in  the eighth condition of Table \ref{z2_table}.    
\begin{table*}[!ht]
\vspace{-10pt}
\caption{Different cases for the data movement for the $Z_n$ and $Z_{n+1}$ computations.} 
\vspace{-10pt}
\centering  
\resizebox{12.5cm}{!}{%

    \begin{tabular}{|c|c|c|c|c| }
        \hline
   $\#$    & Condition   &Data movement& $Z_n$&$Z_{n+1}$\\ 
~&~&during 2 & Computation&Computation\\
~&~&consecutive swaps&&\\\hline  
              
1& $i_{n+1} \neq j_n  \&  i_n \neq j_{n+1} $ & $S_n[i_n] \leftrightarrow S_{n}[j_n]$ & $t_n=S_n[i_n]+S_n[j_n]$&  $t_{n+1}=S_n[i_{n+1}]+S_n[j_{n+1}]$\\
~	& $\& j_{n+1} \neq j_n  $                                 & $S_n[i_{n+1}] \leftrightarrow S_{n}[j_{n+1}]$,&~&~\\\hline

2& $i_{n+1} \neq j_n  \&  i_n \neq j_{n+1} $ & $S_n[i_n] \rightarrow S_{n}[i_{n+1}]$ & $t_n=S_n[i_n]+S_n[j_n=j_{n+1}]$&  $t_{n+1}=S_n[i_{n+1}]+S_n[j_{n+1}=j_n]$\\
~	& $\& j_{n+1} = j_n  $                                 & $S_n[i_{n+1}] \rightarrow S_{n}[j_{n+1}=j_n]$,&~&~\\
~	&~                                                                & $S_n[j_{n+1}=j_n] \rightarrow S_{n}[i_n]$,&~&~\\\hline

3& $i_{n+1} \neq j_n  \&  i_n = j_{n+1} $ & $S_n[i_n=j_{n+1}] \rightarrow S_{n}[j_n]$ & $t_n=S_n[i_n=j_{n+1}]+S_n[j_n]$&  $t_{n+1}=S_n[i_{n+1}]+S_n[i_n=j_{n+1}]$\\
~	& $\& j_{n+1} \neq j_n  $                                 & $S_n[j_n] \rightarrow S_{n}[i_{n+1}]$,&~&~\\
~	&~                                                                & $S_n[i_{n+1}] \rightarrow S_n[i_n=j_{n+1}]$ ,&~&~\\\hline

4& $i_{n+1} \neq j_n  \&  i_n = j_{n+1} $ & $S_n[i_n=j_n=j_{n+1}] \rightarrow S_n[i_{n+1}]$ & $t_n=S_n[i_n=j_n=j_{n+1}]+$&  $t_{n+1}=S_n[i_{n+1}]+$\\
~&$j_{n+1} = j_n $                                    & $S_n[i_{n+1}] \rightarrow S_n[i_n=j_n=j_{n+1}]$ &$S_n[i_n=j_n=j_{n+1}]$&$S_n[i_n=j_n=j_{n+1}]$\\\hline

5& $i_{n+1} = j_n  \&  i_n \neq j_{n+1} $ & $S_n[i_n] \rightarrow S_n(j_{n+1}]$ & $t_n=S_n[i_n]+S_n[j_n=i_{n+1}]$&  $t_{n+1}=S_n[j_n=i_{n+1}]+S_n[j_{n+1}]$\\
~	& $\& j_{n+1} \neq j_n  $                                 & $S_n[j_n=i_{n+1}] \rightarrow S_n[i_n]$,&~&~\\
~	&~                                                                & $S_n[j_{n+1}] \rightarrow S_n[j_n=i_{n+1}]$ ,&~&~\\\hline

6& $i_{n+1} = j_n  \&  i_n \neq j_{n+1} $ & $S_n[i_n] \rightarrow S_n[j_n=j_{n+1}=i_{n+1}];                  $ & $t_n=S_n[i_n]+ $&  $t_{n+1}=S_n[j_n=j_{n+1}=i_{n+1}]$\\
~	& $\& j_{n+1} = j_n  $                                 & $S_n[j_n=j_{n+1}=i_{n+1}] \rightarrow S_n[i_n];$&$S_n[j_n=j_{n+1}=i_{n+1}]$&$+S_n[j_n=j_{n+1}=i_{n+1}]$\\\hline

7& $i_{n+1} = j_n  \&  i_n = j_{n+1} $ &           No Swap & $t_n=S_n[i_n=j_{n+1}] $&  $t_{n+1}=S_n[i_{n+1}=j_n]$\\
~	& $\& j_{n+1} \neq j_n  $                        & ~&$+S_n[j_n=i_{n+1}]$&$+S_n[j_{n+1}=i_n]$\\\hline
8& $i_{n+1} = j_n  \&  i_n = j_{n+1} $ &          Impossible &Discarded&  Discarded\\
~& \& $j_{n+1} = j_n $ &         ~ &~& ~\\\hline
       
    \end{tabular}
}
\label{z2_table} 
\vspace{-10pt}
\end{table*}

\vspace{-10pt}
\subsection{Various methods of Swapping of 2 elements}
\vspace{-10pt}
 In a computer program the execution of swapping of two data elements a and b sitting in two memory locations
x and y respectively involves a temporary variable $t$  and requires 3 operations, such as, x $\rightarrow$ t (a goes to t), 
y $\rightarrow$  x (b goes to x) and $t$ $\rightarrow$  y (a goes to y),  in 3 clocks (assuming 1 operation in 1 clock). In a RC4 software program being executed in an Intel pentium processor, the encryption of 1-byte took seven clocks while each swapping operation took 3 clocks \cite{schneier}. In a CPLD based RC4 hardware the implementation of swapping activity took 3 clocks with each of its three operation 
is executed in 1 clock \cite{kundarewic}. The RC4 block is designed for execution in two parallel FPGA coprocessors and 
implemented in a first generation FPGA takes 3 clocks to encrypt 1-byte \cite{tsoi} in which swapping took 
2 clocks, 1 clock for first 2 operations and 1 clock for the third one. Mathews did a RC4 hardware design for 
execution in one processor within 3 clock cycles in which swap was also being done in 2 clocks - its ASIC implementation 
is reported in  \cite{patent:matthews} and an FPGA implementation with one coprocessor, in \cite{dp:math}. 
Kitos  \cite{patent:matthews} used two RAM blocks as two temporary variables and encrypted 1-byte in 3 clocks in 
which two initial clocks are being spent in making preparation for the swap while the actual swap and also the computation 
of the index  $t$ (vide line 7 of Fig. \ref{fig:ksa_prga_fig}B) based on which the random keystream byte is to be fetched are being done in the third clock. Here one should note that final swap picture settles in hardware in the next clock. Had it been that the 
index  $t$ becomes any one of the two swapping indices ($i$ and $j$ of Fig. \ref{fig:ksa_prga_fig}B), the random keystream would be fetched before the swap is truly being effected and then there remains a likelihood in missing the correct random keystream. This is the reason that the RC4 keystream generation should always be executed in a clock after the swap  as proposed in RC4. In \cite{springerlink:one_byte} and \cite{ieee:two_byte} swapping of 2 bytes together is executed in 1 clock following an innovative circuit design which considers eight logical conditions involving inequaity and/or equality of three pairs of variables ($i_{n+1}$, $j_n$), ($j_{n+1}$, $i_n$) and ($j_{n+1}$ and $j_n$). Adopting an additional pipeline architecture two designs are implemented in \cite{ieee:two_byte}. Of the 2 designs, the design 1 processes 2 PRGA bytes in 2 clocks in which the loop unrolled technique is considered only in KSA for 128 clocks  instead 256 clocks and the conventional RC4 looping in PGRA, while the design 2 processes 2 bytes in 1 clock in which  loop unrolled technique is adopted for both KSA and PRGA units. It has been noticed that both the designs avoided to execute  the swapping and generation of PRGA-bytes stream in the same clocks. 
\vspace{-15pt}
\subsection{Data Processing during rising edges of clock cycles}
Two modes of data processing during rising edges of all clock cycles have been proposed in \cite{springerlink:one_byte} and\cite{ieee:two_byte}.  The issues related to 2-bytes-2-clocks mode of data processing is briefly elaborated in Sec. 2.4.1 following the 2-bytes-1-clock one in Sec. 2.4.2.
\vspace{-10pt}
\subsubsection{2-bytes-2-clocks loop unrolled 3-stage pipeline architecture}
\vspace{-10pt}
2-bytes-2-clocks mode of data processing of S. Sen et. al.  \cite{springerlink:one_byte} is a loop unrolled 3 stage pipe-lined architecture where 3 stages are computed during rising edges of 3 different clocks illustrated in Fig \ref{pipeline1_fig}. In $1^{st}$ stage $i$ and $j$ are computed, in $2^{nd}$ stage swap occurred; and at 3rd stage $Z$ is generated. Though $j$ has dependence on incremented $i$, still $i$ and $j$ both are computed at the same clock. From the hardware aspect the incremented $i$ would not be reflected on $j$ computation while $i$ and $j$ both will be computed at same clock instant. The trick comes in $j$ circuit where instead of $j=j+S[i]$ one needs to make a circuit for $j=j+S[i+1]$. 
\vspace{-10pt}
\begin{figure}[!b]
\vspace{-20pt}
\begin{subfigure}{.5\textwidth}
\begin{tikztimingtable}[
    timing/slope=0,         
    timing/coldist=.0pt,     
    xscale=6.4,yscale=3.2, 
    semithick               
  ]
  \scriptsize  & 0{C}                              \\
\extracode
 \begin{pgfonlayer}{background}
\begin{scope}[gray,semitransparent,semithick]
   
    \foreach \x in {1,...,4}
      \draw (\x,0) -- (\x,-5.8);
  \end{scope}
  \node [anchor=south east,inner sep=0pt]
    at (1.7,-0.0) {\tiny{Stage 1}};
  \node [anchor=south east,inner sep=0pt]
    at (2.7,-0.0) {\tiny{Stage 2}};
  \node [anchor=south east,inner sep=0pt]
    at (3.7,-0.0) {\tiny{Stage 3}};

  \node [anchor=south east,inner sep=0pt]
    at (0.7,-0.40) {\tiny{Cycle 1}};
  \node [anchor=south east,inner sep=0pt]
    at (0.7,-1.6) {\tiny{Cycle 2}};
  \node [anchor=south east,inner sep=0pt]
    at (0.7,-2.80) {\tiny{Cycle 3}};
  \node [anchor=south east,inner sep=0pt]
    at (0.7,-4) {\tiny{Cycle 4}};
  \node [anchor=south east,inner sep=0pt]
    at (0.7,-5.2) {\tiny{Cycle 5}};

\draw [fill=SkyBlue, SkyBlue] (1.0,-0.1) rectangle (2.0,-1.4);
  \node [anchor=south east,inner sep=0pt]
    at (2,-0.45) {\tiny $i_1=i_0+1$};
  \node [anchor=south east,inner sep=0pt]
    at (2,-0.70) {\tiny $j_1=j_0+S_0[i_1]$};
  \node [anchor=south east,inner sep=0pt]
    at (2,-0.95) {\tiny $i_2=i_2+1$};
  \node [anchor=south east,inner sep=0pt]
    at (2,-1.20) {\tiny $j_2=j_1+S_1[i_2]$};

\draw [fill=SkyBlue, SkyBlue] (2.,-1.45) rectangle (3,-2.65);
  \node [anchor=south east,inner sep=0pt]
    at (3,-1.75) {\tiny Swap};
  \node [anchor=south east,inner sep=0pt]
    at (3,-2.0) {\tiny $S_0[i_1], S_0[j_1]$};
  \node [anchor=south east,inner sep=0pt]
    at (3,-2.25) {\tiny Swap };
  \node [anchor=south east,inner sep=0pt]
    at (3,-2.50) {\tiny $ S_1[i_2], S_1[j_2]$};

\draw [fill=SkyBlue, SkyBlue] (3,-2.7) rectangle (4,-3.75);
  \node [anchor=south east,inner sep=0pt]
    at (4,-2.95) {\tiny$Z_1=S_1[S_1[i_1]$};
  \node [anchor=south east,inner sep=0pt]
    at (4,-3.20) {\tiny $+S_1[j_1]]$};
  \node [anchor=south east,inner sep=0pt]
    at (4,-3.45) {\tiny$Z_2=S_2[S_2[i_2]$};
  \node [anchor=south east,inner sep=0pt]
    at (4,-3.70) {\tiny $+S_2[j_2]]$};

\draw [fill=YellowGreen, YellowGreen] (1,-2.7) rectangle (2,-3.75);
  \node [anchor=south east,inner sep=0pt]
    at (2,-2.95) {\tiny $i_3=i_2+1$};
  \node [anchor=south east,inner sep=0pt]
    at (2,-3.2) {\tiny $j_3=j_2+S_2[i_3]$};
  \node [anchor=south east,inner sep=0pt]
    at (2,-3.45) {\tiny $i_4=i_3+1$};
  \node [anchor=south east,inner sep=0pt]
    at (2,-3.7) {\tiny $j_4=j_3+S_3[i_2]$};

\draw [fill=YellowGreen,YellowGreen] (2,-3.75) rectangle (3,-4.7);
  \node [anchor=south east,inner sep=0pt]
    at (3,-3.95) {\tiny Swap};
  \node [anchor=south east,inner sep=0pt]
    at (3,-4.2) {\tiny $S_3[i_2], S_2[j_2]$};
  \node [anchor=south east,inner sep=0pt]
    at (3,-4.45) {\tiny Swap };
  \node [anchor=south east,inner sep=0pt]
    at (3,-4.70) {\tiny $ S_3[i_2], S_0[j_2]$};

\draw [fill=YellowGreen, YellowGreen] (3,-4.7) rectangle (4,-5.75);
  \node [anchor=south east,inner sep=0pt]
    at (4,-4.95) {\tiny$Z_1=S_1[S_1[i_1]$};
  \node [anchor=south east,inner sep=0pt]
    at (4,-5.20) {\tiny $+S_1[j_1]]$};
  \node [anchor=south east,inner sep=0pt]
    at (4,-5.45) {\tiny$Z_2=S_2[S_2[i_2]$};
  \node [anchor=south east,inner sep=0pt]
    at (4,-5.70) {\tiny $+S_2[j_2]]$};
 \end{pgfonlayer}
\end{tikztimingtable}%
\caption{2-bytes-2-clocks 3-stage pipeline architecture}
\vspace{-20pt}
\label{pipeline1_fig}
\end{subfigure}
\begin{subfigure}{.5\textwidth}
\begin{tikztimingtable}[
    timing/slope=0,         
    timing/coldist=.0pt,     
    xscale=10.8,yscale=3.2, 
    semithick               
  ]
  \scriptsize  & 0{C}                              \\
\extracode
 \begin{pgfonlayer}{background}
\begin{scope}[gray,semitransparent,semithick]
   
    \foreach \x in {0.2,...,3}
      \draw (\x,0) -- (\x,-5.8);
  \end{scope}
  \node [anchor=south east,inner sep=0pt]
    at (0.7,-0.0) {\tiny{Stage1}};
  \node [anchor=south east,inner sep=0pt]
    at (1.7,-0.0) {\tiny{Stage2}};
  \node [anchor=south east,inner sep=0pt]
    at (0.1,-0.70) {\tiny{Cycle 1}};
  \node [anchor=south east,inner sep=0pt]
    at (0.1,-3.50) {\tiny{Cycle 2}};

\draw [fill=SkyBlue, SkyBlue] (0.2,-0.1) rectangle (1.2,-1.9);
  \node [anchor=south east,inner sep=0pt]
    at (1.2,-0.45) {\tiny $i_1=i_0+1$};
  \node [anchor=south east,inner sep=0pt]
    at (1.2,-0.70) {\tiny $j_1=j_0+S_0[i_1]$};
  \node [anchor=south east,inner sep=0pt]
    at (1.2,-0.95) {\tiny $i_2=i_2+1$};
  \node [anchor=south east,inner sep=0pt]
    at (1.2,-1.20) {\tiny $j_2=j_1+S_1[i_2]$};
  \node [anchor=south east,inner sep=0pt]
    at (1.2,-1.55) {\tiny Swap  $S_0[i_1], S_0[j_1]$};
  \node [anchor=south east,inner sep=0pt]
    at (1.2,-1.80) {\tiny Swap  $ S_1[i_2], S_1[j_2]$};

\draw [fill=SkyBlue, SkyBlue] (1.2,-1.9) rectangle (2.2,-3.1);
  \node [anchor=south east,inner sep=0pt]
    at (2.1,-2.25){\tiny$Z_1=S_1[S_1[i_1]$};
  \node [anchor=south east,inner sep=0pt]
    at (2.1,-2.5){\tiny $+S_1[j_1]]$};
  \node [anchor=south east,inner sep=0pt]
    at (2.1,-2.7) {\tiny$Z_2=S_2[S_2[i_2]$};
  \node [anchor=south east,inner sep=0pt]
    at (2.1,-2.9) {\tiny $+S_2[j_2]]$};

\draw [fill=YellowGreen, YellowGreen] (0.2,-3.2) rectangle (1.2,-5);
  \node [anchor=south east,inner sep=0pt]
    at (1.2,-3.5) {\tiny $i_3=i_2+1$};
  \node [anchor=south east,inner sep=0pt]
    at (1.2,-3.75) {\tiny $j_3=j_2+S_2[i_3]$};
  \node [anchor=south east,inner sep=0pt]
    at (1.2,-4.0) {\tiny $i_4=i_3+1$};
  \node [anchor=south east,inner sep=0pt]
    at (1.2,-4.25) {\tiny $j_4=j_3+S_3[i_2]$};
  \node [anchor=south east,inner sep=0pt]
    at (1.2,-4.5) {\tiny Swap  $S_2[i_3], S_2[j_3]$};
  \node [anchor=south east,inner sep=0pt]
    at (1.2,-4.75) {\tiny Swap  $ S_3[i_4], S_3[j_4]$};

 \node [anchor=south east,inner sep=0pt]
    at (2.1,-1.8) {\tiny{Cycle 2}};
\draw [fill=YellowGreen, YellowGreen] (1.2,-5) rectangle (2.2,-6.2);
 \node [anchor=south east,inner sep=0pt]
   at (2.1,-5.4){\tiny$Z_3=S_3[S_3[i_3]$};
  \node [anchor=south east,inner sep=0pt]
    at (2.1,-5.65){\tiny $+S_3[j_3]]$};
  \node [anchor=south east,inner sep=0pt]
    at (2.1,-5.9) {\tiny$Z_4=S_4[S_4[i_4]$};
  \node [anchor=south east,inner sep=0pt]
    at (2.1,-6.15) {\tiny $+S_4[j_4]]$};
 \end{pgfonlayer}
\end{tikztimingtable}%
\caption{2-bytes-1-clock loop unrolled 2-stage pipeline architecture  }
\vspace{-10pt}
\label{pipeline2_fig}
\end{subfigure}
\end{figure}
\subsubsection{ 2-bytes-1-clock loop unrolled 2-stage pipeline architecture}
\vspace{-10pt}
In 2-bytes per clock loop unrolled architecture \cite{ieee:two_byte}, the activities undertaken 3 pipeline stages stated in (sec. 2.4.1) are accommodated into 2 pipelines stages in order to increase the throughput and are shown in Fig \ref{pipeline2_fig}.  In $1^{st}$ stage the $i$, $j$ and swap,  operations are processed and in $2^{nd}$ stage $Z$s are extracted. Here also $j$ has data dependence on incremented $i$, and swap has dependence on incremented $i$ as well as on updated $j$. The circuit trick is identical with previous one. Instead of swapping $S[i]$, $S[j]$ registers they have used $S[i+1], S[j+S[i+1]]$. 
\vspace{-20pt}
\section{Conceptual Milestones Contributed in Our Designs and implementations of seven designs }
\vspace{-10pt}
Of the five innovative design concepts described in the fourth paragraph of Introduction, the third and fourth ones are related to
 swapping and data processing respectively. The swapping between two S-box elements, which is the core activity in RC4 from  beginning to end, is executed using a MUX-DEMUX combination. The MUX-DEMUX combination surrounds the S-box and the combined unit is named as Storage Block. The Storage Block has been appropriately designed to mutually swap one pair of S-box elements and also its two pairs, both in 1-clock. The swapping of one pair in 1-clock is termed as 1-byte-1-clock swap and two pairs, as 2-bytes-1-clock swap.  In order to increase throughput, RC4 is suitably organized and many such units are separately installed in many coprocessors functioning in parallel. The RC4 with Composite KSA PRGA (CKP) performing as KSA, PKRS or PRGA are installed in the first coprocessor having a Storage Block and a Z-Circuit.  A stand-alone PGRA unit having a Storage Block and a Z-circuit is installed in each of all coprocessors other than the first.  Both the CKP unit and the stand alone PRGA unit along with all the associated components including the Z-circuit can be designed using a mode of data processing of either 1-byte-1-clock or 2-bytes-1-clock. The data processing is undertaken during both the rising and falling edges of clock cycles instead of using only rising edges or falling edge of clocks. It may be noted that the process of MUX-DEMUX based swap, taking place in clocks having one sensitive-edge, be it of one pair (byte-by-byte processing) or or two pairs (processing of 2-bytes together), done in a particular clock, is considered to be complete in the next clock when the swapped data become available through respective Delayed Flip Flops. The introduction of processing data  in clocks having two sensitive-edges (rising and falling edges) makes it possible to completely execute the MUX-DEMUX based swap during  consecutive two sensitive-edges of two consecutive half clocks thereby making the complete swap to happen in 1-clock. Thus by undertaking the MUX-DEMUX based swap in clocks having two sensitive-edges, it has been possible to introduce 1-byte as well as 2-bytes modes of data processing, both in 1-clock and thereby the throughput has been substantially increased by installing suitable multi-stage pipeline architectures in data processing.  In order to understand the effectiveness of various components so designed, studies on their performance characteristics have been undertaken and following this, eight design ideas have been conceived by systematically increasing coprocessors from one to eight in order to find ways to accelerate RC4 as much as possible in a particular embedded hardware platform. 
\par	From the hardware architecture angle, one can see the designs D1, D2 and D3 as the first hardware architecture with one pair of a coprocessor and a Storage Block, while the designs D4 and D5, as the second one with two pairs of coprocessors and Storage Blocks and the designs D6 and D7, as the third one with four pairs of coprocessors and Storage Blocks. The design D1 uses one coprocessor and implements the conventional RC4 with 1-byte-1-clock mode of data processing. The designs D2 and D3 implements RC4 in one co-processor with CKP having PKRS and PRGA; in D2 the 1-byte-1-clock is the mode of data processing, while in D3 the data processing mode is 2-bytes-1-clock. In D4 and D5, RC4 is implemented in two coprocessors; D4 is a 1-byte-1-clock design, while D5 is designed with 2-bytes-1-clock mode of data processing.  In D6 the RC4 is designed in 1-byte-1-clock mode of data processing and in D7 the same, in 2-bytes-1-clock mode and both are implemented in four coprocessors.  With eight coprocessors the design D8 was appropriately simulated but could not be implemented in hardware due to paucity of necessary resources.  
The designs D1 and D2 give a throughput of 1-byte-in-1-clock, D3 and D4, 2-bytes-in-1-clock, D5 and D6, 4-bytes-in-1-clock and D7 provide a throughput of 8-bytes-in-1-clock.  The timing analyses of all the designs considering data processing during rising and falling edges of clocks have also been undertaken separately for 1-byte-1-clock designs as well as for 2-bytes-1-clock designs.
\par	All circuit issues implemented in designs D1, D2, D4 and D6 involve 1-byte-1-clock mode of data processing and are detailed in Sec.3.1 and all the other designs, D3, D5 and D7 are related to 2-bytes-1-clock mode of data processing and are elaborately discussed in Sec.3.2. 
\vspace{-10pt}
\subsection{1-byte-1-clock mode of Data Processing: Related Circuit Issues}
	The various components processing data in 1-byte-1-clock mode that are used in designs D1, D2, D4 and D6 have been described in the following subsections. The 1-byte-1-clock MUX-DEMUX based Storage Block consisting of the 8-bit S-box is detailed in Sec. 3.1.1 along with the technique adopted in actual swapping. The hardware design of a version of RC4 demonstrated in \cite{ieee:two_byte} has processed two consecutive bytes together and achieved a throughput of 2-bytes-in-2-clocks by adopting 3-stage pipeline architecture. The underlying idea of executing a hardware redesign of conventional RC4 in D1 is to see if the innovative ideas of  swapping based on MUX-DEMUX combination followed by 3-stage pipeline architecture processing data in 1-byte-1-clock mode during both rising and falling edges of all clock cycles can give 1-byte-in-1-clock throughput without resorting to data processing of 2-bytes-in-2-clocks. The design D1 is a homogeneous 1-byte-1-clock design where all circuit components are functioning in the same mode. The design D1 is described in all possible details in Sec.3.1.2. The design D2 is a replica of D1 whose separate KSA and PRGA units are amalgamated in CKP that includes PKRS also and the underlying idea of implementing D2 is to see the  effectiveness of the PKRS in scrambling the S-box 1024 times and to estimate the resource-wise, power-wise and statistic-wise advantage of CKP. The design D2 is well described in Sec. 3.1.3 detailing the CKP including its PKRS and PRGA processes. The D4 and D6 are designs based on two and four coprocessors respectively and for such cases the CKP with in-built PKRS and PRGA is always installed in the first coprocessor and a stand-alone PRGA in other coprocessors. Both the CKP unit and also the stand-alone PRGA include the Storage Block and also the Z-circuit. The PKRS scrambles the 1st S-box 1024 times and at intermediate stages of ints scrambling, the 2nd S-box associated with other coprocessor in D4 is formed and also the 2nd, 3rd and 4th S-boxes associated with other three coprocessors in D6.  All the S-boxes are formed during the scrambling of the 1st S-box and all the S-boxes including the 1st one are in decreasing order of randomization. At an instant, one PRGA byte is fetched from respective S-boxes of decreasing order of randomization, meaning that the 1st PRGA byte comes from 1st S-box, 2nd from 2nd so and so forth. The task of coprocessors other that the 1st one is to generate only PRGA bytes in synchronization with the 1st one and for this a stand-alone PRGA circuit similar to that used in D1 and shown in Figs. 6 is associated with each of all the other coprocessors. The issues related to designs D4 and D6 have been described in detail in Sec.3.1.4.  The Z-circuit which is identical for the CKP unit and also for the stand-alone PRGA unit is presented in Sec.3.1.5.  The timing analyses of 1-byte-1-clock designs of D1, D2, D4 and D6 considering data processing during rising and falling edges of clocks are given in Sec.3.1.6. 
\vspace{-15pt}
\subsubsection{1-byte-1-clock Storage Block: Updating the S-Box following a swap}
\vspace{-10pt}
A single swap without using any temporary variable involves two consecutive statements, x = y and y = x indicating simultaneous register-to-register direct data transfer of two swapping elements. In VHDL, it is noted that a DFF (Delayed Flip Flop) is itself inserted in front of both y and x following two consecutive statements, x = y and y = x indicating that two DFFs hold the respective data till the onset of the next clock. Observing this, it is decided to use an appropriate MUX-DEMUX combination for swapping with a belief that the multiple swaps involving two pairs of S-box elements can also be swapped in one clock. The S-box is placed between DEMUX and MUX and each of 256 outputs of DEMUX accesses respective inputs of 256 S-box elements whose 256 respective outputs are connected to respective inputs of MUX through DFFs. The MUX fetches the required data and passes them appropriately to the DEMUX upgrading the S-box instantaneously and the upgraded data are being hold at the appropriate DFFs to make them available at the input of MUX at the beginning of the next clock. 
\par
\begin{figure}[!htB]
\centering
\vspace{-10pt}
\includegraphics[scale=0.24]{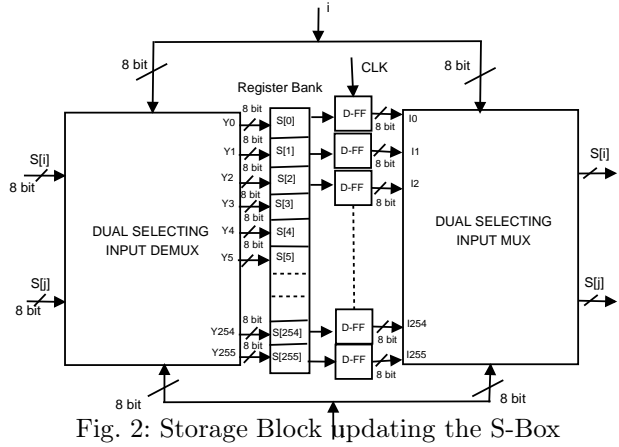}
\vspace{-20pt}
\caption{Storage Block updating the S-Box}
\vspace{-20pt}
\label{fig:s_box}
\end{figure}

 The hardware design of 1-byte-1-clock storage block updating the S-box is shown in Fig. \ref{fig:s_box}. The storage block consists of a register bank containing 256 numbers of 8-bit data representing the S-box, 256:2 MUX, 2:256 DEMUX and 256 D flip-flops (DFFs). The VHDL compiler has merged two 256:1 MUX to design a single 256:2 MUX, the same technique is also adopted for 2:256 DEMUX. Each of the MUX and DEMUX combination is so designed that each accepts 2-select inputs $i$ and $j$ (both of 8 bit width) and addresses the two register data $S[i]$ and $S[j]$ at the same instant. For swaps, the $S[i]$ of MUX is connected to the $S[j]$ of DEMUX and the $S[j]$ of MUX is connected to the $S[i]$ of DEMUX. The storage block has thus 3 input ports ($i$, $j$ and $CLK$), and 2 ports, namely $S[i]$ of MUX \& $S[j]$ of DEMUX and $S[j]$ of MUX \& $S[i]$ of DEMUX.
\subsubsection{1-byte-1-clock design of RC4 with Separate KSA and PRGA Circuits and its implementation : Design D1 }
\label{design1}
\begin{figure}[!ht]
\centering
\vspace{-6pt}
\includegraphics[scale=0.25]{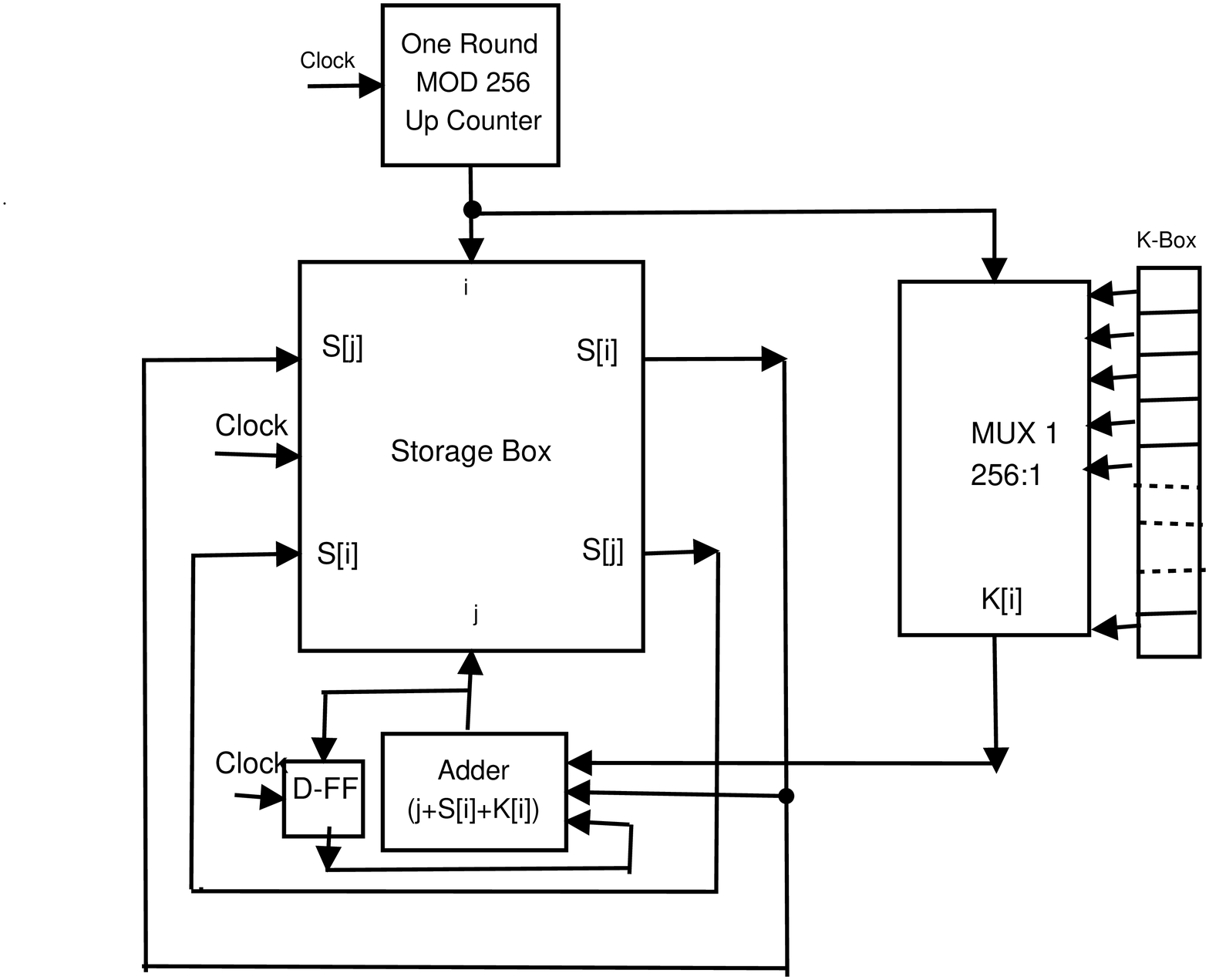}
\vspace{-10pt}
\caption{Schematic Design of 1-byte-1-clock KSA unit}
\vspace{-10pt}
\label{fig:ksa}
\end{figure}

\begin{figure}[!htb]
\centering
\vspace{-10pt}
\includegraphics[scale=0.25]{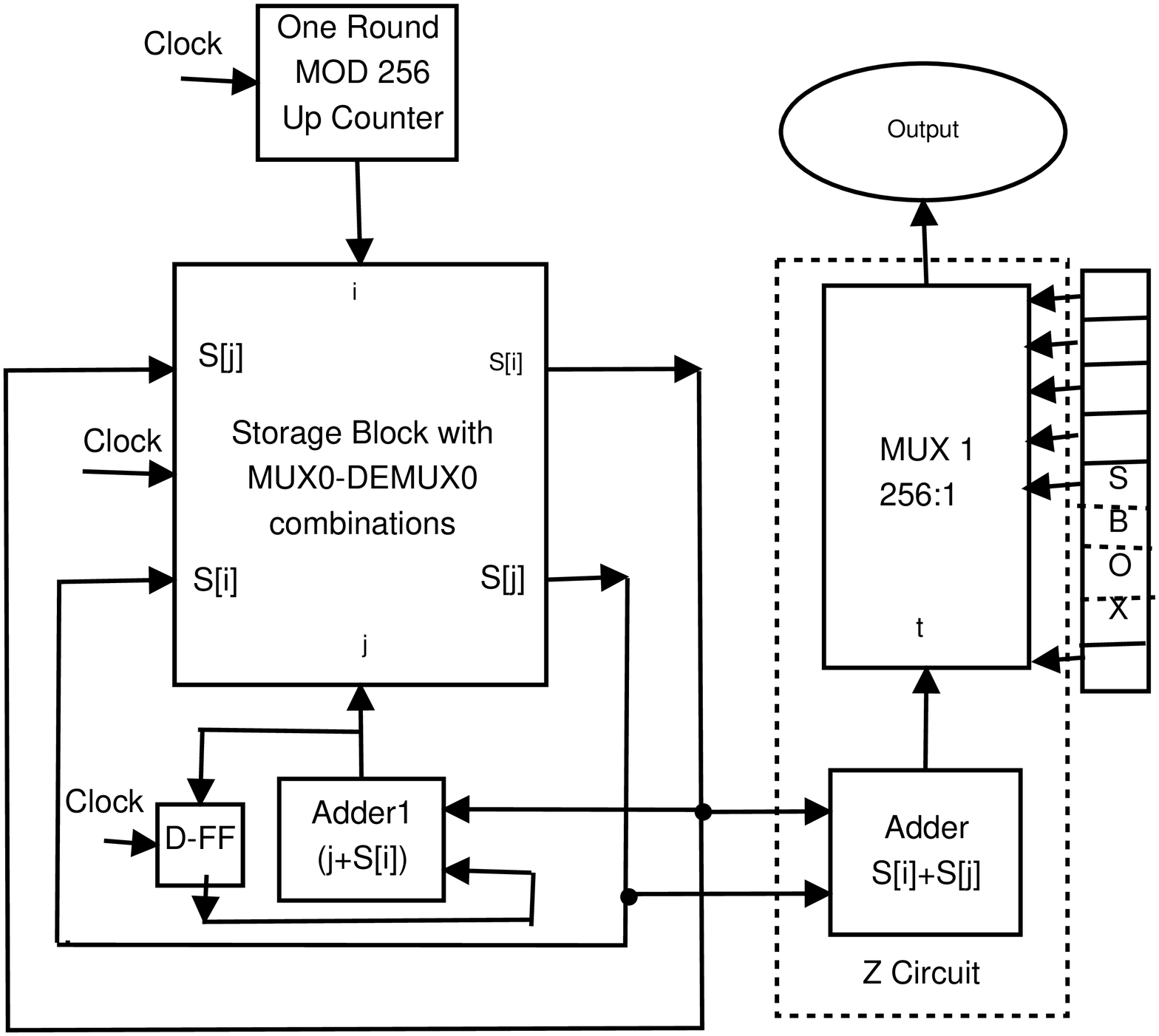}
\vspace{-10pt}
\caption{Schematic Design of 1-byte-1-clock PRGA unit
}
\vspace{-16pt}
\label{fig:prga}
\end{figure}

	The central idea of the present implementation of 1-byte-1-clock conventional RC4 in embedded system is to design its KSA and PRGA units in two separate silicon areas and each of the both uses the identical 1-byte-1-clock storage block shown in Fig. \ref{fig:s_box}.\\
\textbf{Design of the KSA Unit following the algorithmic flow (vide Fig. \ref{fig:ksa_prga_fig}A):} Fig. \ref{fig:ksa} shows a schematic design of the KSA unit. It has a storage block as stated in Sec.3.1.1 whose MUX-DEMUX combination shown in Fig. \ref{fig:s_box} is taken as the MUX0-DEMUX0 combination. Initially the S-Box is filled with identity elements of $i$ whose values change from 0 to 255 as stated in line 3 of initialization module. The l-bytes of secret key are stored in the $K[256]$ array in a repetitive manner as given in line 4 (vide Fig. \ref{fig:ksa_prga_fig}A). The KSA unit does access its storage block with $i$ being provided by a one round of MOD 256 up-counter, providing fixed 256 clock pulses and $j$ being provided by a 3-input adder ($j$, $S[i]$, and $K[i]$) following the line 8 (vide Fig. \ref{fig:ksa_prga_fig}A) where $j$ is clock driven, $S[i]$ is chosen from the S-box being driven by MUX0 of the Storage Block and $K[i]$ is chosen from the K-array and is MUX1 driven (vide Fig.\ref{fig:ksa}).  The S-Box is scrambled by the swapping operation stated in line 9 (vide Fig. \ref{fig:ksa_prga_fig}A) using MUX0-DEMUX0 combination in its Storage Block. The KSA operation takes one initial clock for its initialization and subsequent 256 clock cycles for its execution.
	The strategy of continuity of operations, between the KSA and PRGA processes functioning in two different circuits implemented in two different silicon areas, is to copy the KSA S-box data to the PRGA S-box at the end of the KSA process. \\
\textbf{Design of the PRGA Unit following the algorithmic flow (vide Fig. \ref{fig:ksa_prga_fig}B):}
	Fig.\ref{fig:prga} shows a schematic diagram of the design of the PRGA unit. The PRGA unit does access the storage block having MUX2-DEMUX2 combination with $i$ being provided by a MOD 256 up-counter (vide line 4 of Fig. \ref{fig:ksa_prga_fig}B) and $j$ being given by a 2-input ($j$ and $S[i]$) adder following the line 5 (vide Fig. 1B) where $j$ is clock driven and $S[i]$ is chosen from the S-Box and driven by MUX2 of the Storage Block. With updated $j$ and current $i$, the swapping of $S[i]$ and $S[j]$ is executed following line 6 (vide Fig. \ref{fig:ksa_prga_fig}B) using MUX2-DEMUX2 combination of the storage block in Fig. \ref{fig:prga}. Following the line 7 (vide Fig. \ref{fig:ksa_prga_fig}B), the adder output of $S[i$] and $S[j]$ gives a value of $t$ based on which the key stream $Z$ is selected from the S-Box using MUX3 (vide line 8 of Fig. \ref{fig:ksa_prga_fig}B).

\subsubsection{1-byte-1-clock design of RC4 with CKP functioning as KSA, PKRS and PRGA and its implementation in a coprocessor: Design D2}
\label{design2}
\begin{figure}[!ht]
\centering
\vspace{-10pt}
\includegraphics[scale=0.31]{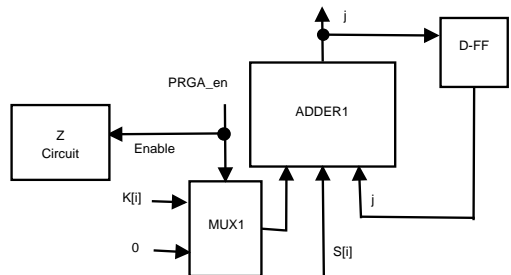}
\vspace{-10pt}
\caption{Design idea of 1-byte-1-clock CKP}
\label{1byte_ckp}
\end{figure}

\begin{figure}[htp!]
\centering
\vspace{-10pt}
\includegraphics[scale=0.2]{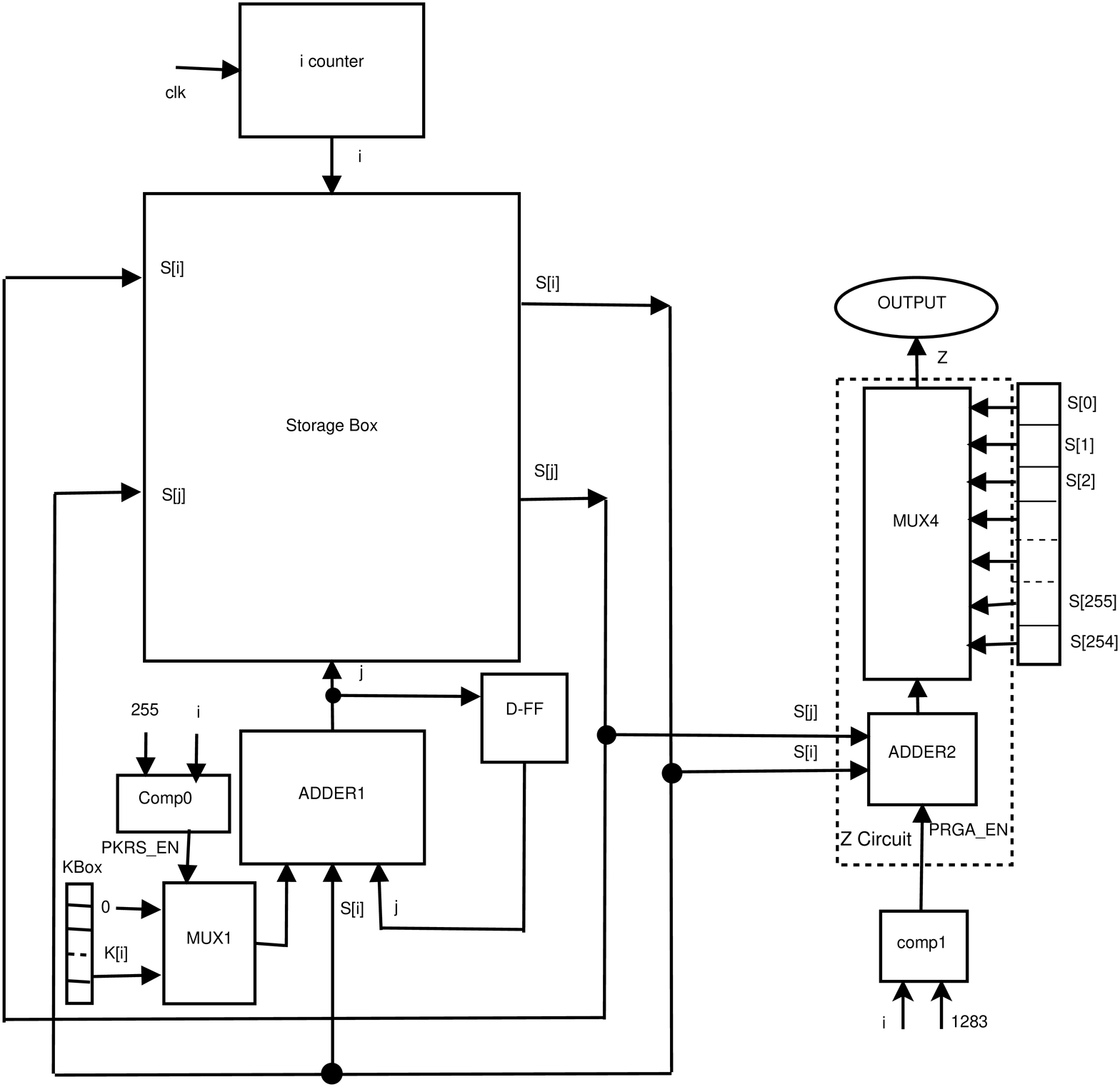}
\vspace{-10pt}
\caption{1-byte-1-clock CKP with PKRS for D2}
\label{1byte_dynamic}
\vspace{-20pt}
\end{figure}

	While designing the 1-byte-1-clock CKP circuit performing the roles of KSA, PKRS and PRGA, one has to sequentially increase $i_n$ and based on the increased $i_{n}$ one finds $j_{n}$ and then $S[i_{n}]$ is swapped with $S[j_{n}]$ in one loop. In RC4, $i$ is considered as the sequential index (line 7 of Fig.1A and line 4 of Fig. 1B) while $j$ is the random index (line 8 of Fig. 1A and line 5 of Fig. 1B) and swapping (line 9 of Fig.1A and line 6 of Fig.1B) between two elements of the S-box is the continuous activity that all processes in CKP undertake.  The operations of CKP for line 8 of KSA ($j_n = j_{n-1} + S[i_n] + K[i_n]$) and for line 5 of PRGA ($j_n=j_{n-1}+S[i_n]$) indicate that the additional K-part in KSA can be arranged in a circuit by MUX1 as shown in Fig.\ref{1byte_ckp}. The other hardware details are shown in Fig.\ref{1byte_dynamic}.  The selecting input signal PKRS\_EN to the MUX is the output terminal of a comparator circuit, comp0 that is looking for the instant when i-count (counting clock cycles) becomes 256 so that the PKRS\_EN status can decide the optional pass of $K$.  When KSA is on, i-count is '0' and the PKRS\_EN latches to '0' after executing the necessary initialization activities during the first clock and the MUX1 passes a Key element (K) to the adder circuit. When the i-count is 257 and KSA is being run for 256 clocks, PKRS\_EN gets enabled and after necessary initialization activity related to the PKRS process being executed during the $258^{th}$ clock gets latched to '1' and '0' is passes to the adder circuit through the MUX1 setting in the PKRS swapping for the next 1024 clocks without any key element. After completion of 1024 cycles of the PKRS process, i-count becomes 1282 and after due initialization activities in regard to the PRGA process during $1283^{rd}$ clock count, the PKRS\_EN is kept in active state in order that identical swapping on account of PRGA continues and a PRGA\_EN terminal which is the output terminal of comparator circuit, comp1 is additionally enabled activating the Z-Circuit to generate PGRA bytes. It may be noted that mutual swapping of two S-box elements does takes place continuously in RC4 from its beginning to the end - in KSA one key element has always a role to choose one of the two swapping elements, in PKRS as well as in PRGA no key element has any role for such a choice and the PRGA process, besides swapping, has a role to activate the Z-circuit and to generate PRGA stream bytes.

\par Fig.\ref{1byte_dynamic} shows the 1-byte-1-clock CKP unit which on activation continues swapping through the processes of KSA, PKRS and PRGA. For the designs D1 and D2 the respective s-boxes undergo identical 256 times of KSA process, but the S-box of D2 undergoes an additional 1024 times of PKRS process that makes the sequence of PRGA bytes obtained from its S-box different to that obtained for D1, more randomized than that of  D1 and free from key bias enunciated by Roos \cite{roos1}
and others  \cite{DBLP:spaul} \cite{springerlink:gpaul}. Looking intricately to the PRGA bytes generation processes in D1 and D2, one would notice the both exhibit identical timing analysis of byte generation processes. 

 \subsubsection{1-byte-1-clock design of RC4 with CKP Circuit together with PKRS and PRGA in the first of coprocessor and Stand-alone PRGA units in other coprocessors: Designs D4 and D6}
\begin{figure}[pb]
\centering
\vspace{-20pt}
\includegraphics[scale=0.38]{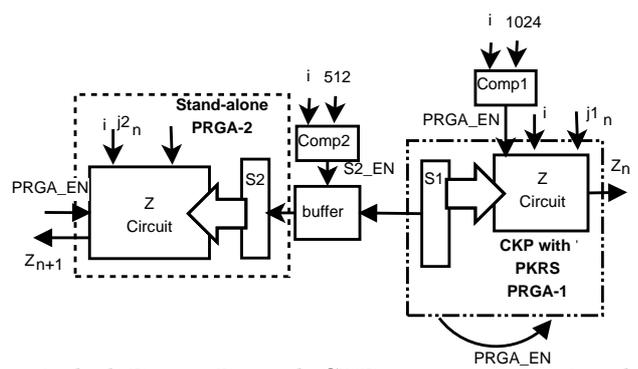}
\vspace{-12pt}
\caption{1-byte-1-clock Design D4 with CKP, in coprocessor-1 and Stand-alone PRGA in coprocessor-2}
\label{2pkrs_ckp1}
\end{figure}

\begin{figure}[!htb]
\centering
\vspace{-20pt}
\includegraphics[scale=0.38]{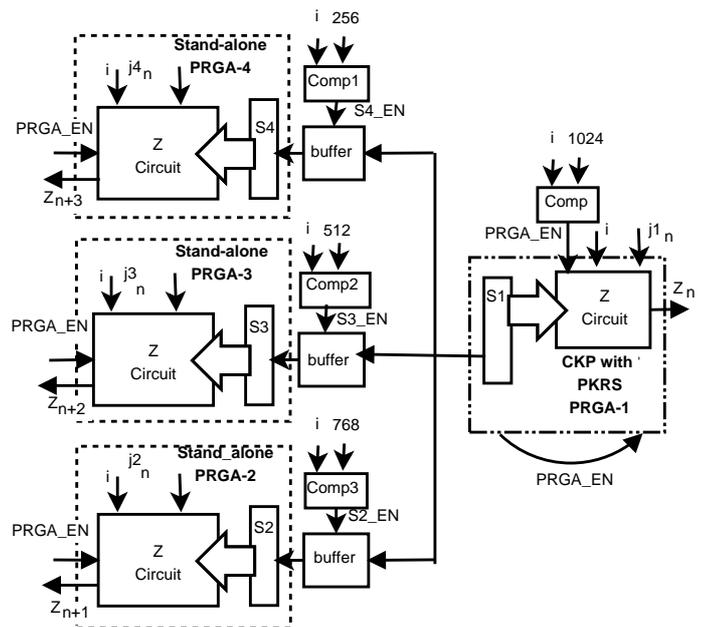}
\vspace{-10pt}
\caption{1-byte-1-clock Design D6 with CKP in coprocessor-1 and stand-alone PRGA in other 3 coprocessors.  }
\label{4pkrs_ckp1}
\end{figure}

\textbf{Design D4:} A 1-byte-1-clock accelerated RC4 is designed in D4 involving two coprocessors, two S-Boxes and two Z-circuits -- it generates two Zs in 1-clock.  The CKP unit that has already been designed in D2 involving one coprocessor and shown in Fig.\ref{2pkrs_ckp1} is installed in the  $1^{st}$ coprocessor with some additional part in the PKRS process and a stand-alone PRGA unit identical to that designed in D1 as a separate PRGA unit and shown in Fig.6 is installed in the  $2^{nd}$ coprocessor.  The two RC4 units in D4 each have a Storage Block with an identity S-box and a Z-circuit. The architecture of the broad hardware design of D4 is shown in Fig.\ref{2pkrs_ckp1} where the additional PKRS part in-built with its CKP is explicitly shown over and above the corresponding one used in D2. The KSA process in D4 is identical to that used in D2 and continues scrambling the $1^{st}$ S-box for 256 cycles with key element and at the end of KSA PKRS\_EN gets enabled and sets in the PKRS process - this aspect has not been shown in Fig.\ref{2pkrs_ckp1}. The PKRS part in D4 continues scrambling the 1st S-box ($S_1$) for 1024 clocks, but while doing so it has an additional task of forming its 2nd S-box ($S_2$) in the 2nd coprocessor out of S1 at an intermediate stage of its scrambling.  The comparator circuit, comp2 looking for the instant when clock counter completes i-count of 512, and the S2\_EN is instantly enabled activating the buffer through which the $1^{st}$ S-box at the $513^{rd}$ instant is copied to the $2^{nd}$ S-box and after this S2\_EN is disabled.  The comparator circuit, comp1 after the PKRS process completing 1024 clock cycles, PRGA\_EN is enabled activating together the Z-circuit of the 1st processor and the stand-alone PRGA unit including its own Z-circuit of $2^{nd}$ coprocessor keeping PKRS\_EN in active state in order that identical swapping on account of PRGA continues. In the design D4, the two S-boxes in two coprocessors together contribute a pair of PRGA bytes in its generated sequence. Of each pair of bytes, the $1^{st}$ byte is from the  $1^{st}$  S-box and the $2^{nd}$ one, from the $2^{nd}$ S-box. In order to keep a count of Zs, the 2 S-boxes, $S_1$ and $S_2$, together consider identical sequential index $i$ and two different random indices $j1$ and $j2$ for $S_1$ and $S_2$ respectively and contribute 1 PRGA byte each in 1-clock making D4 to produce a throughput of 2-bytes-in-1-clock.\\
\textbf{Design D6:} Considering four coprocessors and four S-boxes, another 1-byte-1-clock version of accelerated RC4 generating 4 Zs in 1-clock is designed in D6 and is shown in Fig.\ref{4pkrs_ckp1}.  The CKP unit together with PKRS and PRGA
 that has been  designed in D4 is installed in the 1st coprocessor of D6 with some additional part in PKRS and a stand-alone PRGA unit that is identical to that used in D4 is installed in three coprocessors of D6.  The four RC4 units in D6 each have a Storage Block with an identity S-box and a Z-circuit. The design D6 in closely similar to the design D4 with an additional task for PKRS in D6 is to form three S-boxes out of the scrambling S1 S-box at three different stages of its randomization. The KSA process in CKP here is identical to that used in D2 as well as in D4 and continues scrambling the 1st S-box (S1) with key elements for 256 clock cycles. As soon as the KSA process of the CKP unit ends after 256 clocks, the PKRS\_EN is activated setting in the PKRS process. When the PKRS process runs for 256 clocks, S4\_EN is activated and data of S1 is copied to S4 through a buffer; when the PKRS continues for another 256 clocks, the S3\_EN is activated making ways to copy S1 data to S3; when the PKRS continues for another 256 clocks, the S2\_EN is enabled and S1 data are copied to S2 through a buffer; when PKRS runs for another 256 clocks, it finishes 1024 clocks of run and then the PKRS\_EN is kept in active state in order that identical swapping on account of PRGA continues and PRGA\_EN is activated enabling the Z circuits associated with the 1st coprocessors and the three stand-alone PRGA circuits associated with three other coprocessors including the respective Z-circuits allowing each coprocessor to consider the identical sequential index $i$ and four different random indices j1, j2, j3 and j4 for S1, S2, S3 and S4 respectively and to generate 1 PRGA bytes in 1-clock, thereby D6 produces 4-bytes in 1-clock. 
\subsubsection{1-byte-1-clock Z-Circuits Generating Random Key Streams}
In this mode of data processing the Z-circuit generating random key stream is same for all the designs: (1) in D1, separate PRGA unit shown in Fig.\ref{fig:prga} generates Z, (2) in D2, the CKP coprocessor shown in Fig.\ref{1byte_dynamic} generates Z, (3) in D4 and D6, the $1^{st}$ coprocessor with CKP shown in Fig.8 generates Z, and (4) one stand-alone PRGA unit, in D4 and three stand-alone PRGA units in D6 generates Z. For all the cases mentioned above the Z-circuit is identical having two inputs to an adder circuit from two inout ports of the Storage Block having an S-box with a MUX-DEMUX combination. The first inout port is the S[i] of MUX and S[j] of DEMUX and the second inout port is the S[j] of MUX and S[i] of DEMUX. The 't' in the adder output terminal indicates an 8-bit index using which the MUX3 in Fig.\ref{fig:prga} and MUX4 in Fig.\ref{1byte_dynamic}, picks up the appropriate 8-bit data from the S-box and sends the same to the output of the respective MUX as 'Z' being a data in the sequence of random key stream.
\subsubsection{Timing analyses of 1-byte-1-clock Designs: D1, D2, D4 and D6}
\begin{figure}
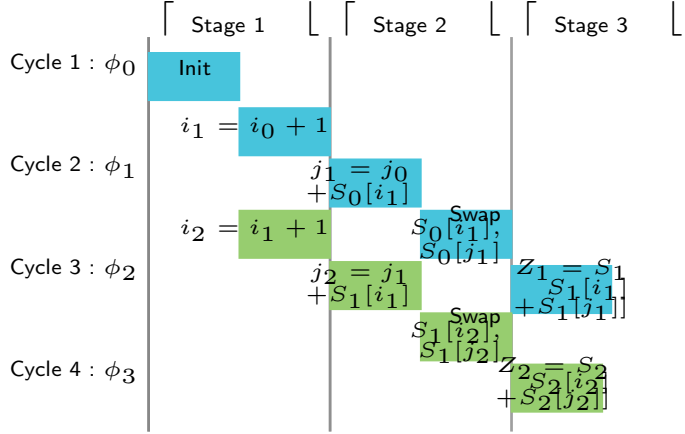

\resizebox{9cm}{!}{%
\begin{tikztimingtable}[
    timing/slope=0,         
    timing/coldist=.0pt,     
    xscale=7.3,yscale=3.2, 
    semithick               
  ]
  \scriptsize  & 0{C}                              \\
\extracode
 \begin{pgfonlayer}{background}
\begin{scope}[gray,semitransparent,semithick]
   
    \foreach \x in {0.25,...,4}
      \draw (\x,0) -- (\x,-5);

  \end{scope}
  \node [anchor=south east,inner sep=0pt]
    at (0.4,-0.0) {$ \lceil$ };
  \node [anchor=south east,inner sep=0pt]
    at (.9,-0.0) {\tiny{Stage 1}};
  \node [anchor=south east,inner sep=0pt]
    at (1.2,-0.0) {$ \lfloor$ };
  \node [anchor=south east,inner sep=0pt]
    at (1.4,-0.0) {$ \lceil$ };
  \node [anchor=south east,inner sep=0pt]
    at (1.9,-0.0) {\tiny{Stage 2}};
  \node [anchor=south east,inner sep=0pt]
    at (2.2,-0.0) {$ \lfloor$ };
  \node [anchor=south east,inner sep=0pt]
    at (2.4,-0.0) {$ \lceil$ };
  \node [anchor=south east,inner sep=0pt]
    at (2.9,-0.0) {\tiny{Stage 3}};
  \node [anchor=south east,inner sep=0pt]
    at (3.2,-0.0) {$ \lfloor$ };

  \node [anchor=south east,inner sep=0pt]
    at (0.2,-0.50) {\tiny{Cycle 1 : $\phi_0$}};
  \node [anchor=south east,inner sep=0pt]
    at (0.2,-1.80) {\tiny{Cycle 2 : $\phi_1$}};
  \node [anchor=south east,inner sep=0pt]
    at (0.2,-3.1) {\tiny{Cycle 3 : $\phi_2$}};
  \node [anchor=south east,inner sep=0pt]
    at (0.2,-4.4) {\tiny{Cycle 4 : $\phi_3$}};

\draw [fill=SkyBlue, SkyBlue] (.25,-0.2) rectangle (.75,-0.8);
  \node [anchor=south east,inner sep=0pt]
    at (0.6,-0.5) {\tiny Init};

\draw [fill=SkyBlue, SkyBlue] (0.75,-0.9) rectangle (1.25,-1.5);
  \node [anchor=south east,inner sep=0pt]
    at (1.25,-1.3) {\tiny $i_1=i_0+1$};

\draw [fill=SkyBlue, SkyBlue] (1.25,-1.55) rectangle (1.75,-2.15);
  \node [anchor=south east,inner sep=0pt]
    at (1.7,-1.85) {\tiny $j_1=j_0$};
  \node [anchor=south east,inner sep=0pt]
    at (1.7,-2.15) {\tiny $+S_0[i_1]$};

;
\draw [fill=SkyBlue, SkyBlue] (1.75,-2.2) rectangle (2.25,-2.8);
  \node [anchor=south east,inner sep=0pt]
    at (2.2,-2.4) {\tiny Swap};
  \node [anchor=south east,inner sep=0pt]
    at (2.2,-2.6) {\tiny $S_0[i_1],$};
  \node [anchor=south east,inner sep=0pt]
    at (2.2,-2.9) {\tiny $ S_0[j_1]$};

\draw [fill=SkyBlue, SkyBlue] (2.25,-2.9) rectangle (2.8,-3.5);
  \node [anchor=south east,inner sep=0pt]
    at (2.9,-3.10) {\tiny$Z_1=S_1$};
  \node [anchor=south east,inner sep=0pt]
    at (2.9,-3.35) {\tiny$S_1[i_1]$};
  \node [anchor=south east,inner sep=0pt]
    at (2.9,-3.6) {\tiny $+S_1[j_1]]$};



\draw [fill=YellowGreen, YellowGreen] (0.75,-2.2) rectangle (1.25,-2.8);
  \node [anchor=south east,inner sep=0pt]
    at (1.25,-2.6) {\tiny $i_2=i_1+1$};

\draw [fill=YellowGreen, YellowGreen] (1.25,-2.85) rectangle (1.75,-3.45);
  \node [anchor=south east,inner sep=0pt]
    at (1.7,-3.15) {\tiny $j_2=j_1$};
  \node [anchor=south east,inner sep=0pt]
    at (1.7,-3.45) {\tiny $+S_1[i_1]$};

\draw [fill=YellowGreen, YellowGreen] (1.75,-3.5) rectangle (2.25,-4.1);
  \node [anchor=south east,inner sep=0pt]
    at (2.2,-3.7) {\tiny Swap};
  \node [anchor=south east,inner sep=0pt]
    at (2.2,-3.9) {\tiny $S_1[i_2],$};
  \node [anchor=south east,inner sep=0pt]
    at (2.2,-4.15) {\tiny $ S_1[j_2]$};

\draw [fill=YellowGreen, YellowGreen] (2.25,-4.15) rectangle (2.75,-4.75);
  \node [anchor=south east,inner sep=0pt]
    at (2.8,-4.35) {\tiny$Z_2=S_2$};
  \node [anchor=south east,inner sep=0pt]
    at (2.8,-4.55) {\tiny$S_2[i_2]$};
  \node [anchor=south east,inner sep=0pt]
    at (2.8,-4.75) {\tiny $+S_2[j_2]]$};
 \end{pgfonlayer}
\end{tikztimingtable}%
}
\caption{Proposed 1 byte per Clock}
\vspace{-20pt}
\label{pipeline1_usfig}
\end{figure}

	Both the designs D1 and D2 are 1-byte-1-clock designs in one coprocessor with one S-box and the difference between D1 and D2 lies in pre-PRGA activities, in the sense that only KSA process is being run in D1 for 257 clock cycles while in D2 PKRS process is being run for 1025 clock cycles after KSA as it is in D1.  The PRGA process in D1 starts at the onset of 257th clock cycle while the same in D2 starts from 1283rd clock cycle.  It may be noted that the sequence of PRGA bytes of both are different to each other, but the time sequences of PRGA activities for both being the same give identical timing analysis. 
One may note that the in both the architectures of \cite{springerlink:one_byte} and \cite{ieee:two_byte} the combinational logic load in each clock is quite large that is likely to lead to the increase of critical path. The logic load in each stage may be reduced by increasing the number of pipeline stages.  In the present paper, with an eye to reduce the critical path delay, it is proposed to introduce four pipeline stages by suitably reducing logic loads in each stage of state as well as space and in order to increase the speed of execution it is also proposed to undertake data processing activity in the pipeline stages during both the clock edges, instead of one of the two clock edges. In $1^{st}$ stage only $i$ values are incremented, in $2^{nd}$ stage $j$ values are updated, in $3^{rd}$ stage swap occurred, and in final stage Zs are generated. It may be noted that the dual clock edge sensitive circuits may not be a very good option for circuits with heavy combinational loads - however, the same are not true for circuits with light combinational loads. \\
\textbf{Data Flow Timing in D2 (PRGA Unit only)}\\
The proposed dual clock edge sensitive 1-byte 1-clock loop unrolled architecture in 4-pipeline stages is shown in Fig.10. The MOD 256 up counter shown in Fig.6 is so designed that $i$ starts from '1', goes up to '255' and then it repeats from '0' to '255' for each 256 subsequent clock cycles.  The dual-clock-edge sensitive timing analysis of D2 for the first four clock cycles is shown below. All additions shown in the timing analysis are modulo 256 additions.
\begin{enumerate}
\item Rising edge of  $\phi_0$: $j_0$=0; $i_0$=0.
\item Falling edge of $\phi_0$: $i_1$ =1.
\item Rising edge of  $\phi_1$: $j_1$=($j_0$ + $S[i_1]$).
\item Falling edge of $\phi_1$: $i_2 \rightarrow$2;  S[$i_1$]  $\leftrightarrow S[j_1]$;
\item Rising edge of  $\phi_2: j_2=(j_1+S[i_2]);$  $Z_1=(S[i_1]+S[j_1])$.
\item Falling edge of $\phi_2$: $i_3 \rightarrow$3;  S[$i_2$] $\leftrightarrow$S[$j_2$]; 
\item Rising edge of $\phi_3$: $j_3=(j_2+S[i_3])$; $Z_2=(S[i_2]+S[j_2])$. 
\end{enumerate}
The series continues generating successive random key streams (Zs). If the text characters are n, $Z_n$ will be executed after (n+2) PRGA clocks. Its throughput per byte is (1+2/n). The timing diagram is shown in Fig.\ref{pipeline1_usfig}.

\textbf{Data Flow Timing in D4 (PRGA units only)}
The D4 has two coprocessors, the 1st one is the identical one installed in D2 with little modification in PKRS and the $2^{nd}$ one is the stand-clone PRGA unit. Two separate S-boxes, S1 and S2, are separately attached with two coprocessors. In order to keep a count of Zs, the 2 S-boxes, S1 and S2, together consider identical sequential index i since it is derived from the clock and two different random indices j1 and j2 as local for S1 and S2 respectively.  At an instant the two S-boxes contribute a pair of byte-stream, thereby $1^{st}$ S-box contributes all odd streams, namely $Z_1$, $Z_3$. $Z_5$ etc. while the $2^{nd}$ one, to the even streams i.e. $Z_2$, $Z_4$, $Z_6$ etc. The dual-clock-edge sensitive timing analysis of D4 for the first four clock cycles is also shown below.  The 1st line in each clock edge belongs to the  $1^{st}$ S-box coupled with $1^{st}$ coprocessor, while the $2^{nd}$ line, to the $2^{nd}$ S-box.  
\begin{enumerate}
\item Rising edge of  $\phi_0$ : $i_0$=0e $j_{10}$=0,. $j_{20}$=0. 
\item Falling edge of $\phi_0$ :  $i_1 \rightarrow$ 1.
\item Rising edge of  $\phi_1$ : $j1_{1}=(j1_{0} + S_1[i_1])$; $j2_{1}=(j2_{0} + S_2[i_1])$.
\item Falling edge of $\phi_1$ : $i_2 \rightarrow 2 ; S_1[i_1] \leftrightarrow S[j1_{1}] ; S_2[i_2] \leftrightarrow S[j2_{1]}$;
\item Rising edge of  $\phi_2$ : $j1_{2} = (j1_{1} + S_1[i_2]); Z_{11} = (S[i_1] + S[j_{11}])  = Z_1.  j22 = (j21 +  S2[i2]);  Z21 = (S[i1] + S[j21]) \% 256 = Z_2$ .
\item Falling edge of $\phi_2$ : $i_3 \rightarrow 3; S1[i_1] \leftrightarrow S[j1_{2}] ; S_2[i_2] \leftrightarrow S[j2_{2}]$ ;
\item Rising edge of  $\phi_3$ :$ j_{13} $ = ($j_{12}$ + S2[i3]); Z12 = ($S[i_2]$ + $S[j1_2]$) = Z3;  $j2_3$ = ($j2_2$ + $S_2[i_3])$;  Z22 = ($S[i_2]$ + S[$j1_2$])  = Z4.
\end{enumerate}
In design D6 one may note that in each clock edges of all the clock cycles the $3^{rd}$ and $4^{th}$ coprocessors contribute 1 PRGA byte each after a lapse of 2 clocks over and above the same contributed by $1^{st}$ and $2^{nd}$ coprocessors. 
\subsection{Circuit Issues related to 2-bytes-1-clock modes of Design}
The crux of 2-bytes-1-clock modes of RC4 design is to simultaneously upgrade $i_n$ \& $i_{n+1}$ from $i_{n-1}$ and $j_n$ \& $j_{n+1}$ from $j_{n-1}$ and to simultaneously execute swapping of $S[i_n]$ with $S[j_n]$ and of $S[i_{n+1}]$ with $S[j_{n+1}]$ in one storage block installed in a coprocessor functioning as 2-bytes-1-clock CKP having built-in PKRS and PRGA and also in another storage block installed in another coprocessor functioning parallel as stand-alone 2-bytes-1-clock PRGA. The upgradation of two sequential indices $i_n$ and $i_{n+1}$ simultaneously from $i_{n-1}$ is simple and is executed by using an adder adding always '1' with $i_{n-1}$ and by using another adder adding always '2' with $i_{n-1}$. The simultaneous upgradation of two random indices $j_n$ \& $j_{n+1}$ from $j_{n-1}$ is rather complex and the same in 2-bytes-1-clock CKP and also in 2-bytes-1-clock stand-alone PRGA are described in Sec.3.2.1 and 3.2.2 respectively with sub-section headings $j_n$ and $j_{n+1}$ generator. The MUX-DEMUX based swap of 2 pair of bytes together is executed by a 2-bytes-1-clock Storage Block which has been described in Sec.3.2.3.  For simultaneous execution of two swaps, it is necessary to look into the seven conditions stated in Table 1 and after knowing actual bytes undergoing swap operation the swapping of necessary bytes would be undertaken. A behavior model based on the said seven conditions has been developed using VHDL and the unit is named as swap controller which is described in Sec.3.2.4.        
\par The D3 is a RC4 design with 2-bytes-1-clock CKP with in-built PKRS and PRGA implemented in one coprocessor and is discussed in Sec.3.2.5 describing the PKRS process in detail.  It may be noted that the mutual arrangement of data elements in D3 S-box after each swap of 2 pair of bytes together becomes identical to that in D2 after the corresponding 2 sequential swaps. Hence, if the PKRS process scrambles the D3 S-box for 512 times, it would achieve the level of randomization identical to that obtained in D2 by running the PKRS for 1024 times and then D3 produces bytes sequence identical to that produced by D2.  The D5 and D7 are designs of RC4 involving 2 and 4 coprocessors respectively and for such cases the 2-bytes-1-clock CKP with in-built PKRS and PRGA like the one in D3 is always installed in the 1st coprocessor and a stand-alone 2-bytes-1-clock PRGA unit, in other coprocessors. The issues related to designs D5 and D7 have been described in detail in Sec.3.2.6.  The Z-circuits associated with the CKP unit installed in the $1^{st}$ coprocessor and also associated with the stand-alone PGRA unit installed in other coprocessors are the same and each of all coprocessors generates 2-PRGA bytes together in 1-clock from respective post-PKRS S-boxes. The design of the Z-circuit is described in Sec. 3.2.7. The timing analyses of 2-bytes-1-clock mode of data processing during rising and falling edges of clock cycles designed in suitable pipeline architectures for 2-bytes-1-clock designs of D3, D5 and D7 are presented in Sec.3.2.8. 
\vspace{-10pt}
\subsubsection{$j_n$ and $j_{n+1}$ Generator from $j_{n-1}$ for 2-bytes-1-clock CKP Unit}
\begin{figure}[!htbp]
\centering
\vspace{-6pt}
\includegraphics[scale=.26]{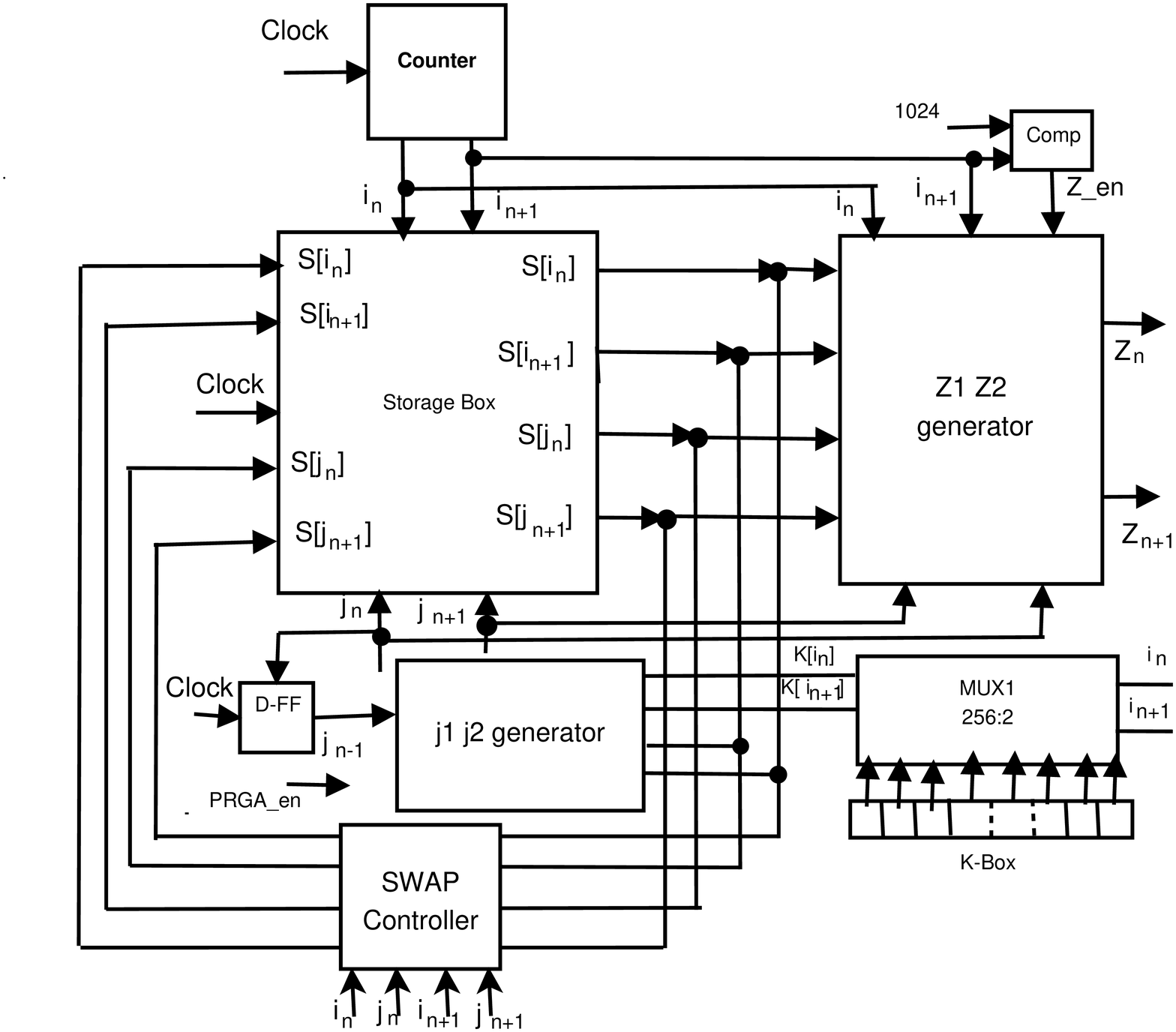}
\vspace{-6pt}
\caption{CKP unit of 2 byte per Clock Hardware}
\vspace{-10pt}
\label{d3:ckt}
\end{figure}

\begin{figure}[!htbp]
\centering
\vspace{-6pt}
\includegraphics[scale=0.27]{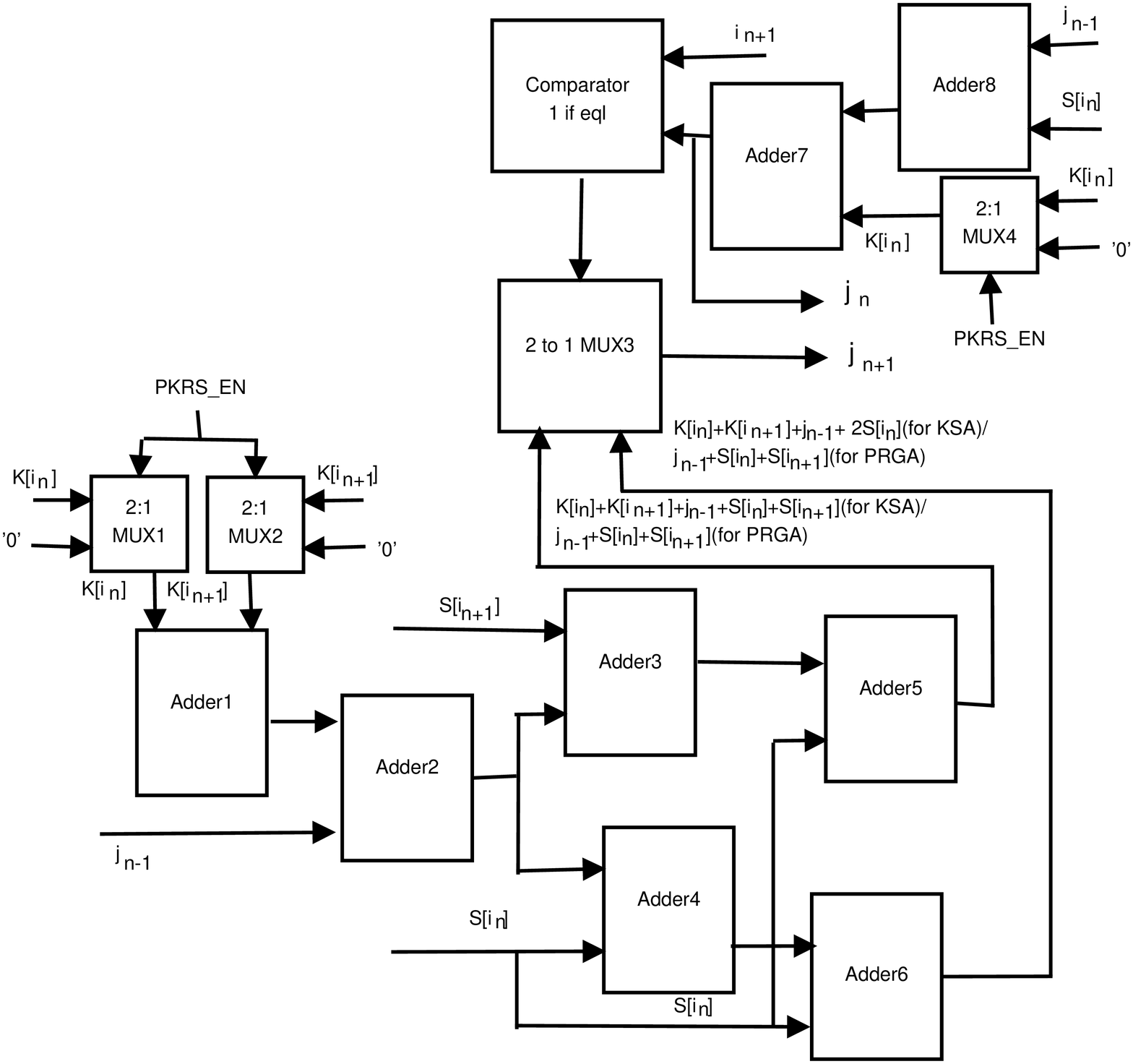}
\vspace{-15pt}
\caption{2-bytes-1-clock CKP generating $j_n$ and $j_{n+1}$ from $j_{n-1}$ }
\vspace{-5pt}
\label{fig:dkp_j1_j2}
\end{figure}

Fig.\ref{d3:ckt} shows the 2-byte-1-clock 1st coprocessor with CKP generating $j_n$  \& $j_{n+1}$ from $j_{n-1}$ with the help of a swap controller. As stated earlier the sequential indices $i_n$  and $i_{n+1}$ are simultaneously upgraded from $i_{n-1}$ using two adders and the $S[i_n]$ and $S[i_{n+1}]$ being simultaneously fetched from the S-box are fed to its MUX unit. One has to adopt the identical circuit techniques while fetching $K[i_n]$ and $K[i_{n+1}$] from the K-box. In Fig. \ref{fig:dkp_j1_j2} the simultaneous up-gradations of $j_n$ and $j_{n+1}$ from $j_{n-1}$ in CKP are shown.  In the upper part of Fig. \ref{fig:dkp_j1_j2}, $j_n$ is shown upgraded from $j_{n-1}$ by adding $j_{n-1}$ with $S[i_{n-1}+1]$ in Adder8 whose result is added in Adder7 with $K[i_{n-1}+1]$ chosen using MUX1 and the result of Adder7 is $j_n$. For evaluation of $j_{n+1}$ directly from $j_{n-1}$, both $K[i_{n-1}+1]$ and $K[i_{n-1}+2]$ are chosen using MUX2 and MUX3 respectively and are added in Adder1 as shown in Fig.  \ref{fig:dkp_j1_j2}. The result of Adder1 is added with $j_{n-1}$ in Adder2 whose result is added with $S[i_{n-1}+2]$ in Adder3 and also with $S[i_{n-1}+1]$ in Adder4; the result of Adder3 is further added with $S[i_{n-1}+1]$ in Adder5 and the result of Adder4 is also added with $S[i_{n-1}+1]$ in Adder6, as shown in Fig.\ref{fig:dkp_j1_j2}. The results of Adder5 and Adder6 are $j_{n+1}$ for two logical conditions stated in eq.(3).  The comparator shown in Fig.12 dynamically chooses one of the two logical conditions between $i_{n+1}$ and jn and the right $j_{n+1}$ is dynamically chosen using MUX4. It may be noted that after completion of 128 clocks the PKRS\_EN is enabled and it, instead of passing of $K[i_{n-1}+1]$ and $K[i_{n-1}+2]$, passes two '0's to Adder1 and also instead of passing of $K[i_{n-1}+1]$ passes '0' to Adder8 and the same thereby activates the PKRS process for further 512 clocks followed by continuation of the PRGA process.
\vspace{-10pt}
\subsubsection{$j_n$ and $j_{n+1}$ Generator from $j_{n-1}$ for 2-bytes-1-clock Stand-alone PRGA Unit} 
\begin{figure}[!htb]
\centering
\vspace{-6pt}
\includegraphics[scale=0.32]{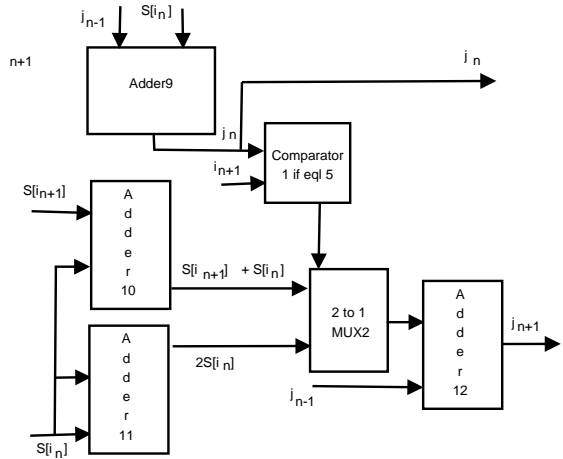}
\vspace{-10pt}
\caption{2-bytes-1-clock stand-alone PRGA generating $j_n$ and $j_{n+1}$ from $j_{n-1}$ }
\vspace{-10pt}
\label{fig:j2prga}
\end{figure}

For 2-bytes-1-clock Stand-alone PRGA Unit, the simultaneous upgradation of in and $i_{n+1}$ from $i_{n-1}$ involves a different circuit, since no key bytes are considered from the K-box. Fig.\ref{fig:j2prga} shows the necessary circuit where $j_n$ is obtained by adding $j_{n-1}$ with $S[i_{n-1}+1]$ in Adder9. The Adder10 adds $S[i_{n-1}+2]$ with $S[i_{n-1}+1]$ and Adder11 adds $S[i_{n-1}+1]$ with itself and one of the two outputs from Adder10 and Adder11 is dynamically chosen using MUX5 based on the logical condition between $i_{n+1}$ and $j_n$ provided by the comparator, as given in eq.(3) and is added with $j_{n-1}$ in Adder12, the output of which is the dynamic $j_{n+1}$.  
\vspace{-10pt}
\subsubsection{2-bytes-1-clock Storage Block: Updating the S-Box following a swap}
\begin{figure}[!htb]
\centering
\vspace{-10pt}
\includegraphics[scale=0.26]{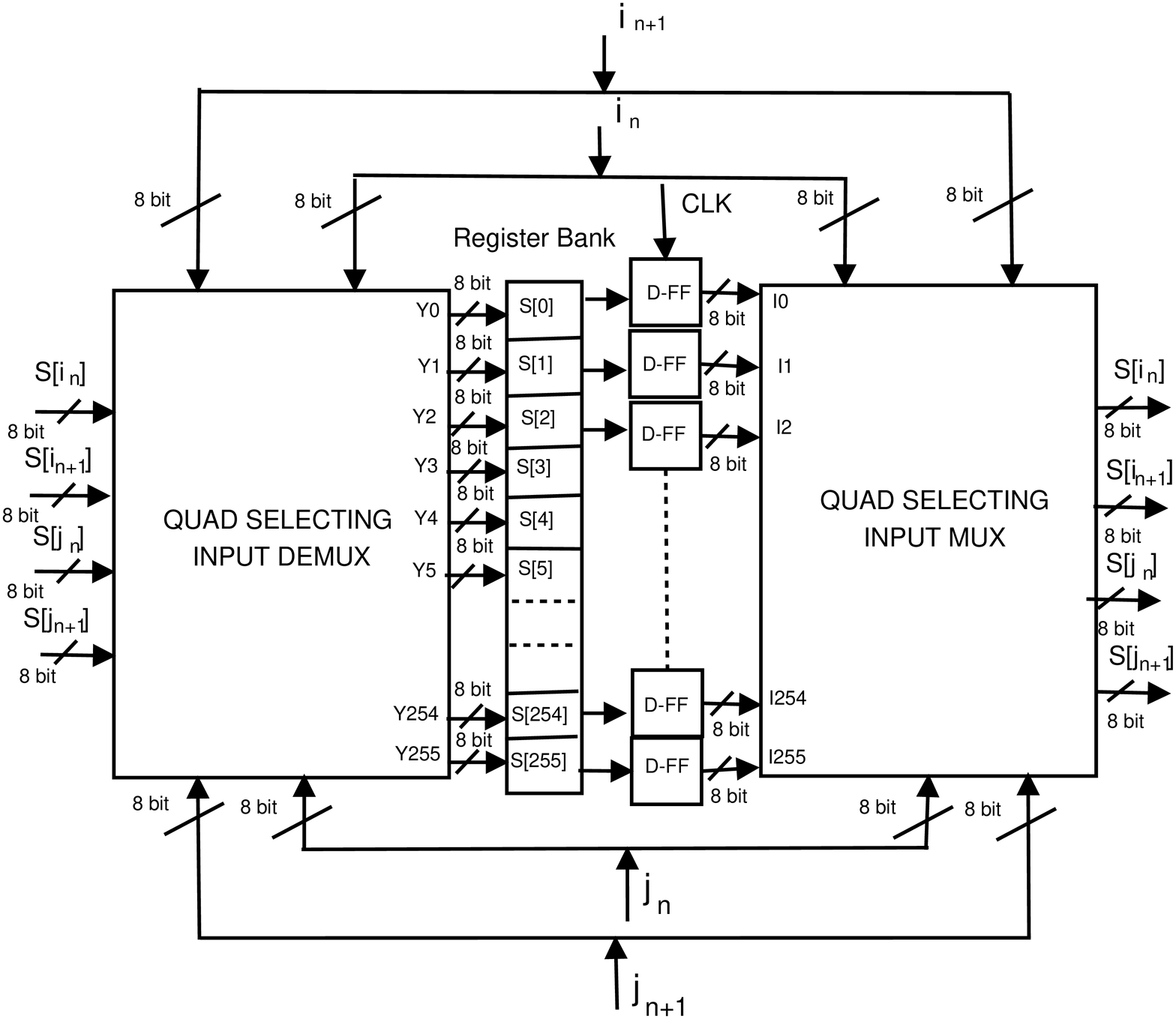}
\vspace{-6pt}
\caption{Storage Block updating the S-Box}
\vspace{-10pt}
\label{fig:s_box2}
\end{figure}

Fig. \ref{fig:s_box2} shows a schematic diagram of the design of the 2-bytes-1-clock storage box unit based on MUX-DEMUX combination that functions in the identical way the 1-byte-1-clock Storage Block shown in Fig. 4 does function. The difference lies in number of ports of the respective S-boxes. This storage block also consists of a register bank containing 256 numbers of 8-bit data representing the S-Box (register bank), 256:1 MUX, 1:256 DEMUX and 256 D Flip-Flops. Here two sets of $i$ and $j$ can address 4 S-box elements at a time. The values of $S[i_n]$, $S[j_n]$ and $S[i_{n+1}]$, $S[j_{n+1}]$ are updated by a 4 input DEMUX followed by a 4 input MUX through a Swap Controlling block, explained in Sec.3.2.2.

\subsubsection{2-bytes-in-1-clock Swap Controller}
 While swapping of 2 pair of bytes in 2-bytes-1-clock storage block using MUX-DEMUX combination, its MUX unit picks up 4 bytes, $S[i_n]$, $S[j_n]$, $S[i_{n+1}]$ and $S[j_{n+1}]$ from an S-box and passes them to the swap controller in which four controlling indices $i_n$, $j_n$, $i_{n+1}$ and $j_{n+1}$ are additionally fed. Considering the seven conditions of data transfer cases depicted in columns 1 and 2 of Table \ref{z2_table}, one can design a behavioral model of the swap controller which has 4 controlling input ports to be received from $i_n$, $j_n$, $i_{n+1}$ and $j_{n+1}$ and 4 input ports receiving data $S[i_n]$, $S[j_n]$, $S[i_{n+1}]$ and $S[j_{n+1}]$ from the MUX and has 4 output ports receiving the swapped data to be fed to the DEMUX. The Pictorial presentation of the swap controller is also depicted on Fig. \ref{d3:ckt}. Dynamically the four controlling input variables would always satisfy one of the three set of logical conditions listed in the first seven rows of the "Condition" column of Table 1 and the swap controller would follow the swapping data shown in the corresponding row of the first seven rows depicted in the "Data Movement during 2 swaps" column of the same Table and appropriately connects the 4 input data lines to the 4 output ports to be fed to the DEMUX. The storage block of PRGA unit provides 2 output ports from its 2 in ports which are fed to an adder circuit with MUX3. During the falling edge of a clock pulse, $S[i]$ and $S[j]$ values corresponding to ith and jth locations of the register bank are read and put on hold to the respective D-Flip-Flops. During the rising edge of the next clock pulse, the $S[i]$ and $S[j]$ values are transferred to the MUX outputs and instantly passed to the $S[j]$ and $S[i]$ ports of the DEMUX respectively and in turn are written to the jth and ith locations of the register bank. The updated S-Box is ready during the next falling edge of the same clock pulse.

\subsubsection{2-bytes-1-clock design of RC4 with CKP functioning as KSA, PKRS andPRGA and its implementation in a coprocessor: Design D3}
\label{design3}
Fig.  \ref{d3:ckt} shows the design layout of the 2-bytes-1-clock CKP circuit employing related KSA, PKRS and PRGA processes.  If 2-bytes together are processed in 1-clock by adopting a suitable CKP circuits, it is necessary to decide whether the KSA will be run for 256 clocks or for 128 clocks. In the present study of various designs, it is intended that the introduction of PKRS should have identical effects to the Post-PKRS PRGA byte sequences of all designs. If the 2-bytes-1-clock KSA runs for 128 clocks, the post-PKRS PRGA-bytes sequence to be obtained in D3 would be identical to that obtained in design D2. If 2-bytes-1-clock KSA unit runs for 256 clocks, its PKRS unit receives a randomized S-box which the 1-byte-1-clock PKRS unit would have received had it been that its KSA unit is being run for 512 clocks. For the present case of 2-bytes-1-clock design, it is decided to run KSA for 128 clocks instead of 256 and PKRS for 512 clocks instead of 1024 clocks. 
Only the loop unrolled approach reduces the number of iterations of PKRS process from 1024 to 512. While the KSA is done, PKRS\_EN activates PKRS and undertakes swapping of the S-Box 1024 times in 512 (2 swaps in each round) rounds using 2 sets of $i$ and $j$ indices without enabling the Z circuit. After 512 rounds PRGA\_EN activates the PRGA circuits installed at all the other coprocessors including their Z-circuits and the Z circuit of all the 1st coprocessor so that all coprocessors can generate key stream parallel from respective S-boxes.
\subsubsection{2-bytes-1-clock designs of RC4 with CKP functioning as KSA, PKRS and PRGA  and its implementation in Coprocessors: Designs D5 and D7}
\begin{figure}[!htb]
\centering
\vspace{-10pt}
\includegraphics[scale=0.35]{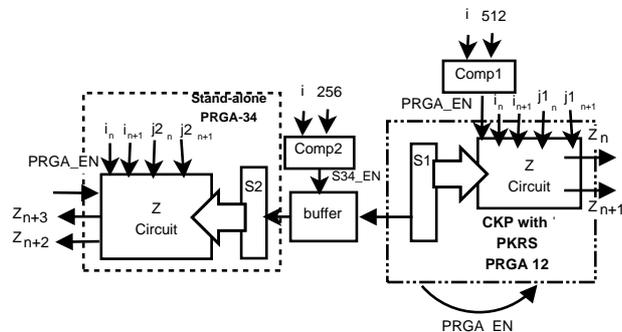}
\vspace{-12pt}
\caption{2-bytes-1-clock CKP with PKRS for 1st coprocessor in D5}
\label{2pkrs_ckp2}
\end{figure}

\begin{figure}[!htb]
\centering
\vspace{-10pt}
\includegraphics[scale=0.37]{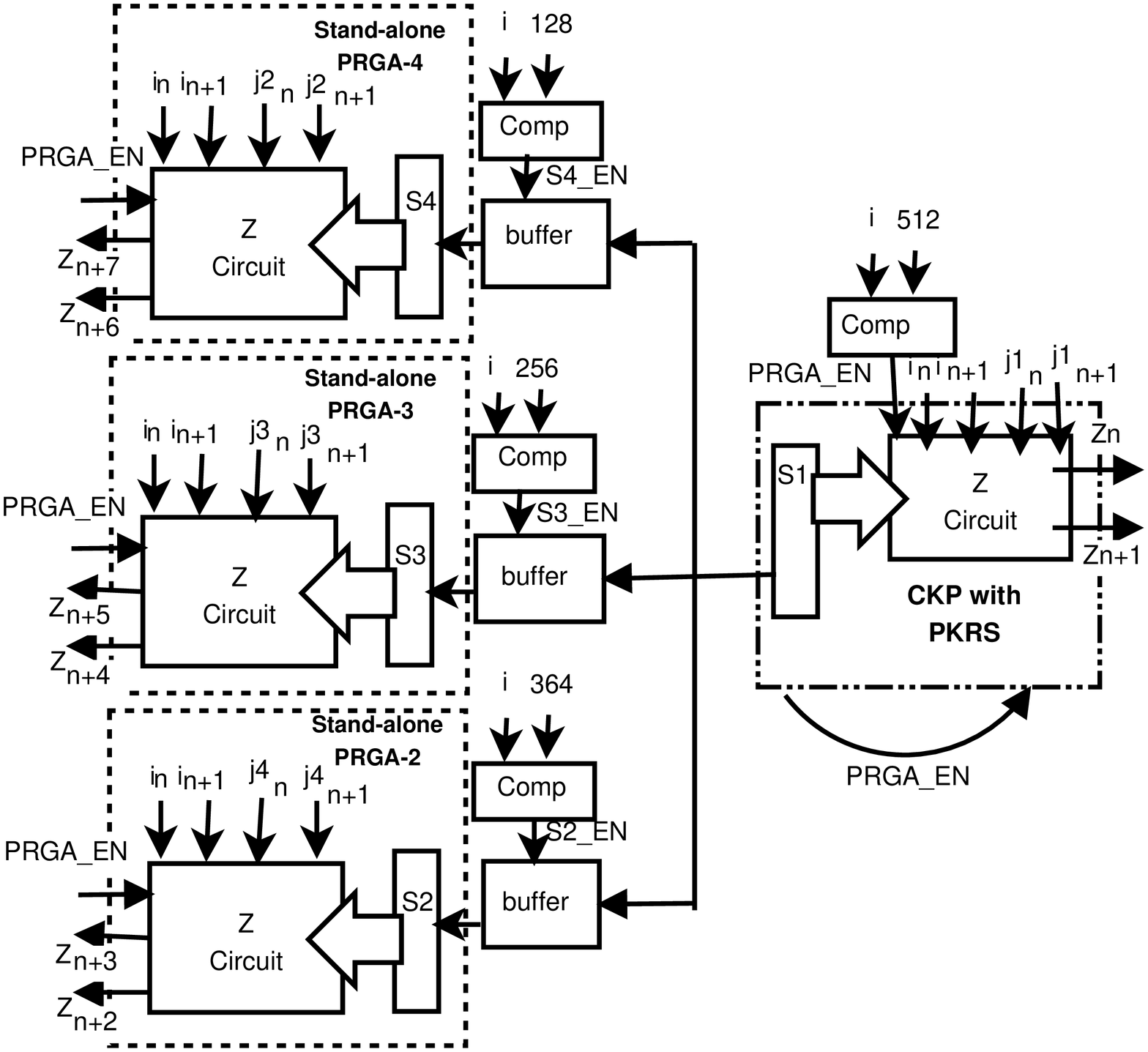}
\vspace{-10pt}
\caption{2-bytes-1-clock CKP in coprocessor-1 and stand-alone PRGA in 3 coprocessors}
\label{4pkrs_ckp2}
\end{figure}

During the PKRS process undertaken in D5 having 2 co-processors and 2 S-Boxes, the S-box, named as S2, is buffered after 256 count of $i$ and the original S-Box, named as S1 is kept after the 512 count of i. D5 produces 4 Zs in 1-clock, 2 from S2 and 2 from S1.
The sequential index $i$ being the clock count is identical for both S1 and s2, while the random indices for S1 and S2 are considered as $j_1$ and $j_2$ respectively.
 The hardware architecture of said design is shown in Fig. \ref{2pkrs_ckp2}. During the PKRS process undertaken in D7 having 4 co-processors and 4 S-Boxes, the S-boxes, named as S4, S3 and S2, are buffered after 128, 256 and 384 counts of $i$ respectively and the original S-Box, named as S1 is considered after the 512 count of i. The i-count in this case also remains the same for the four S-boxes, but the different j-count for S1, S2, S3 and S4 is considered as $j_1$, $j_2$,$j_3$ and $j_4$ respectively.
D7 produces 8 Zs in 1-clock, 2 from each of the 4 post-PKRS S-Boxes. The hardware architecture of said design is shown in Fig. \ref{4pkrs_ckp2}. Make Z\_EN as PRGA\_EN and make necessary changes in 2 j-counts in each S-box. 
\vspace{-10pt}
\vspace{-20pt}
\subsubsection{2-bytes-1-clock Z-Circuit:  Generation of Key Streams $Z_n$ \& $Z_{n+1}$ together}
\begin{figure}[htbp]
\centering
\vspace{-9pt}
\includegraphics[scale=0.27]{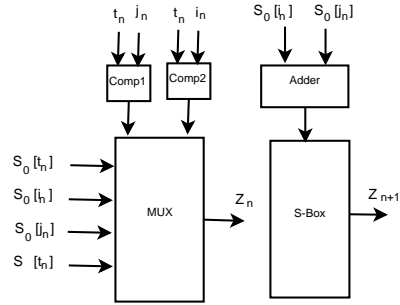}
\vspace{-6pt}
\caption{$Z_n$ \& $Z_{n+1}$ Generator}
\vspace{-8pt}
\label{fig:z1z2}
\end{figure}

In RC4, generation of $Z_n$ takes place from $S_n$ level S-box form after 1 pair of swap in $S_{n-1}$-level S-box. 
In 2-bytes-1-clock mode of design the 2 pairs of swapping brings the S-box from $S_{n-1}$-level directly to $S_{n+1}$-level bypassing the $S_n$-level and  $Z_n$ \& $Z_{n+1}$ are generated together in 1-clock. It is planned to fetch $i_n$ and $j_n$ from  $S_{n-1}$-level S-box before swap and the post-swap $Z_n$ is computed following a simple trick presented in eq. \ref{zn1_equ} and \ref{zn2_equ} below,
\begin{equation}
\label{zn1_equ}
t_{n} = (S_{n-1}[i_{n}]+S_{n-1}[j_{n}]) mod 256	
\end{equation}
\begin{equation}
\label{zn2_equ}
\begin{split}
if ((t_{n}!=i_{n}) and (t_{n}!=j_{n}))~~ Z_{n}=S_{n-1}[t_{n}];\\
	else if ((t_{n}=i_{n}) and (t_{n}!=j_{n})) ~~Z_{n}=S_{n-1}[j_{n}];\\
	else ((t_{n}!=i_{n}) and (t_{n}=j_{n}))~~ Z_{n}=S_{n-1}[i_{n}]; 
\end{split} 
\end{equation}
The $Z_{n+1}$ is computed from $S_{n+1}$-level S-box after the completion of 2-bytes-in-1-clock swap following the usual computation of RC4 done in 1-byte-in-1-clock design.  The circuit computing $Z_n$ fetches $S_{n-1}[i_n]$ and $S_{n-1}[j_n]$ from the input side of the swap controller during the falling clock edge and put them in an adder, the output of which along with $i_n$ and $j_n$ are fed to a MUX through necessary comparators so that the condition stated in eq.5 above can be appropriately organized enabling to fetch suitable data from the $S_{n-1}$-level S-box as dictated by eq. \ref{zn2_equ}. The circuit computing $Z_{n+1}$ picks up $i_{n+1}$ and $j_{n+1}$ from output side of the swap controller during the rising clock edge following the falling clock edge and put them to an adder, the output of which is fed to an MUX which picks up appropriate data from the S2-level S-box. The hardware architecture of $Z_n$ and $Z_{n+1}$ are shown in Fig \ref{fig:z1z2}.
\vspace{-10pt}
\subsubsection{Timing analyses of 2-bytes-1-clock Designs: D3, D5 and D7}

\begin{figure}
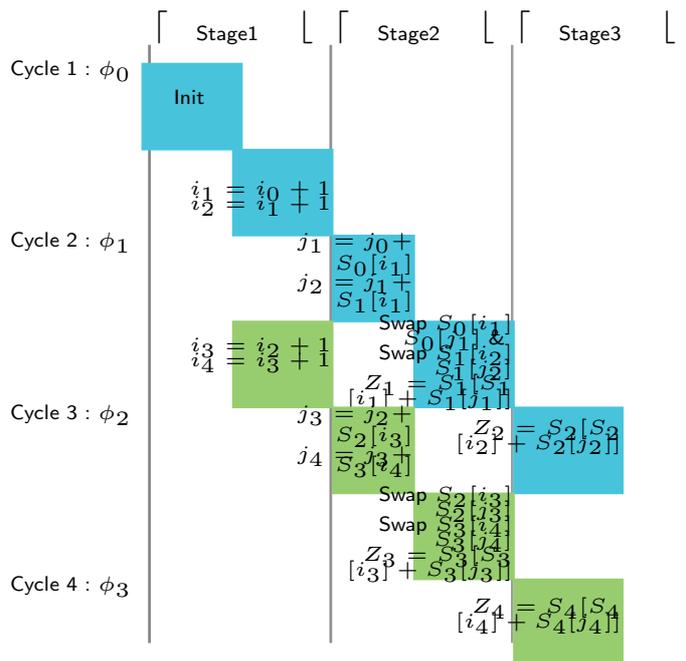

\resizebox{9cm}{!}{%
\begin{tikztimingtable}[
    timing/slope=0,         
    timing/coldist=.0pt,     
    xscale=7.7,yscale=3.2, 
    semithick               
  ]
  \scriptsize  & 0{C}                              \\
\extracode
 \begin{pgfonlayer}{background}
\begin{scope}[gray,semitransparent,semithick]
   
    \foreach \x in {0.29,...,4}
      \draw (\x,0) -- (\x,-8);

  \end{scope}
  \node [anchor=south east,inner sep=0pt]
    at (0.4,-0.0) {$ \lceil$ };
  \node [anchor=south east,inner sep=0pt]
    at (.9,-0.0) {\tiny{Stage1}};
  \node [anchor=south east,inner sep=0pt]
    at (1.2,-0.0) {$ \lfloor$ };
  \node [anchor=south east,inner sep=0pt]
    at (1.4,-0.0) {$ \lceil$ };
  \node [anchor=south east,inner sep=0pt]
    at (1.9,-0.0) {\tiny{Stage2}};
  \node [anchor=south east,inner sep=0pt]
    at (2.2,-0.0) {$ \lfloor$ };
  \node [anchor=south east,inner sep=0pt]
    at (2.4,-0.0) {$ \lceil$ };
  \node [anchor=south east,inner sep=0pt]
    at (2.9,-0.0) {\tiny{Stage3}};
  \node [anchor=south east,inner sep=0pt]
    at (3.2,-0.0) {$ \lfloor$ };

  \node [anchor=south east,inner sep=0pt]
    at (0.2,-0.50) {\tiny{Cycle 1 : $\phi_0$}};
  \node [anchor=south east,inner sep=0pt]
    at (0.2,-2.8) {\tiny{Cycle 2 : $\phi_1$}};
  \node [anchor=south east,inner sep=0pt]
    at (0.2,-5.1) {\tiny{Cycle 3 : $\phi_2$}};
  \node [anchor=south east,inner sep=0pt]
    at (0.2,-7.4) {\tiny{Cycle 4 : $\phi_3$}};

\draw [fill=SkyBlue, SkyBlue] (.25,-0.25) rectangle (.8,-1.4);
  \node [anchor=south east,inner sep=0pt]
    at (0.6,-0.8) {\tiny Init};

\draw [fill=SkyBlue, SkyBlue] (0.75,-1.4) rectangle (1.3,-2.55);
  \node [anchor=south east,inner sep=0pt]
    at (1.3,-2.1) {\tiny $i_1=i_0+1$};
 \node [anchor=south east,inner sep=0pt]
    at (1.3,-2.3) {\tiny $i_2=i_1+1$};

\draw [fill=SkyBlue, SkyBlue] (1.3,-2.55) rectangle (1.75,-3.7);
  \node [anchor=south east,inner sep=0pt]
    at (1.75,-2.8) {\tiny $j_1=j_0+$};
  \node [anchor=south east,inner sep=0pt]
    at (1.75,-3.1) {\tiny $S_0[i_1]$};
  \node [anchor=south east,inner sep=0pt]
    at (1.75,-3.35) {\tiny $j_2=j_1+$};
  \node [anchor=south east,inner sep=0pt]
    at (1.75,-3.6) {\tiny $S_1[i_1]$};

\draw [fill=SkyBlue, SkyBlue] (1.75,-3.7) rectangle (2.3,-4.85);
  \node [anchor=south east,inner sep=0pt]
    at (2.3,-3.9) {\tiny Swap $S_0[i_1]$};
  \node [anchor=south east,inner sep=0pt]
    at (2.3,-4.1) {\tiny $ S_0[j_1]$ \& $$};
  \node [anchor=south east,inner sep=0pt]
    at (2.3,-4.3) {\tiny Swap $S_1[i_2]$};
  \node [anchor=south east,inner sep=0pt]
    at (2.3,-4.5) {\tiny $ S_1[j_2]$};
  \node [anchor=south east,inner sep=0pt]
    at (2.3,-4.7) {\tiny$Z_1=S_1[S_1$};
  \node [anchor=south east,inner sep=0pt]
 at (2.3,-4.9) {\tiny $[i_1]+S_1[j_1]]$};

\draw [fill=SkyBlue, SkyBlue] (2.3,-4.85) rectangle (2.9,-6);
  \node [anchor=south east,inner sep=0pt]
    at (2.9,-5.3) {\tiny$Z_2=S_2[S_2$};
  \node [anchor=south east,inner sep=0pt]
    at (2.9,-5.5) {\tiny $[i_2]+S_2[j_2]]$};



\draw [fill=YellowGreen, YellowGreen] (0.75,-3.7) rectangle (1.3,-4.85);
  \node [anchor=south east,inner sep=0pt]
    at (1.3,-4.2) {\tiny $i_3=i_2+1$};
 \node [anchor=south east,inner sep=0pt]
    at (1.3,-4.4) {\tiny $i_4=i_3+1$};

\draw [fill=YellowGreen, YellowGreen] (1.3,-4.85) rectangle (1.75,-6);
  \node [anchor=south east,inner sep=0pt]
    at (1.75,-5.1) {\tiny $j_3=j_2+$};
  \node [anchor=south east,inner sep=0pt]
    at (1.75,-5.4) {\tiny $S_2[i_3]$};
  \node [anchor=south east,inner sep=0pt]
    at (1.75,-5.65) {\tiny $j_4=j_3+$};
  \node [anchor=south east,inner sep=0pt]
    at (1.75,-5.8) {\tiny $S_3[i_4]$};

\draw [fill=YellowGreen, YellowGreen] (1.75,-6) rectangle (2.3,-7.15);
  \node [anchor=south east,inner sep=0pt]
    at (2.3,-6.2) {\tiny Swap $S_2[i_3]$};
  \node [anchor=south east,inner sep=0pt]
    at (2.3,-6.4) {\tiny $ S_2[j_3]$};
  \node [anchor=south east,inner sep=0pt]
    at (2.3,-6.6) {\tiny Swap $S_3[i_4]$};
  \node [anchor=south east,inner sep=0pt]
    at (2.3,-6.8) {\tiny $ S_3[j_4]$};
  \node [anchor=south east,inner sep=0pt]
  at (2.3,-7) {\tiny$Z_3=S_3[S_3$};
  \node [anchor=south east,inner sep=0pt]
    at (2.3,-7.2) {\tiny $[i_3]+S_3[j_3]]$};

\draw [fill=YellowGreen, YellowGreen] (2.3,-7.15) rectangle (2.9,-8.3);
  \node [anchor=south east,inner sep=0pt]
    at (2.9,-7.7) {\tiny$Z_4=S_4[S_4$};
  \node [anchor=south east,inner sep=0pt]
    at (2.9,-7.9) {\tiny $[i_4]+S_4[j_4]]$};

 \end{pgfonlayer}
\end{tikztimingtable}%
}
\vspace{-10pt}
\caption{Proposed 2 byte per Clock}
\vspace{-10pt}
\label{pipeline2_usfig}
\end{figure}

\begin{enumerate}
\item Rising edge of  $\phi_0$: $j_0=0$; $i_0$=0.
\item Falling edge of $\phi_0$: $i_1$=1; $i_2$=2.
\item Rising edge of  $\phi_1$: $j_1$=$(j_0 + S_0[i_1])$; $j_2=j_0+S_0[i_1]+S_1[i_2]$.
\item Falling edge of $\phi_1$: $i_3=$3; and $i_4=4$, Swap Occurred, $Z_1=S_1(S_0[i_1]+S_0[j_1])$;
\item Rising edge of  $\phi_2: j_3=(j_2+S_2[i_3]);$ $j_4=(j_2+S_2[i_3]+S_3[i_4]);$  $Z_2 = S_2[S_1[i_2] + S_1[j_2]]$.
\item Falling edge of $\phi_2$: $i_5=$5; $i_6=6$, Swap Occurred, $Z_3=(S_2[i_3]+S_2[j_3])$; 
\item Rising edge of  $\phi_3: j_5=(j_4+S_4[i_5]);$ $j_6=(j_4+S_2[i_5]+S_5[i_6]),$,  $Z_4=S_3[S_2[i_4]+S_2[j_4]$.
\end{enumerate}
It is to be noted that the timing diagram of PKRS is slimier to to PRGA except the $Z$ computation. The timing diagram is shown  in fig.\ref{pipeline2_usfig}.  
\vspace{-15pt}
\section{Results of all Implementations : Comparative Study}
\label{ram}
\begin{figure}[ht]
\centering
\vspace{-10pt}
\includegraphics[scale=0.185]{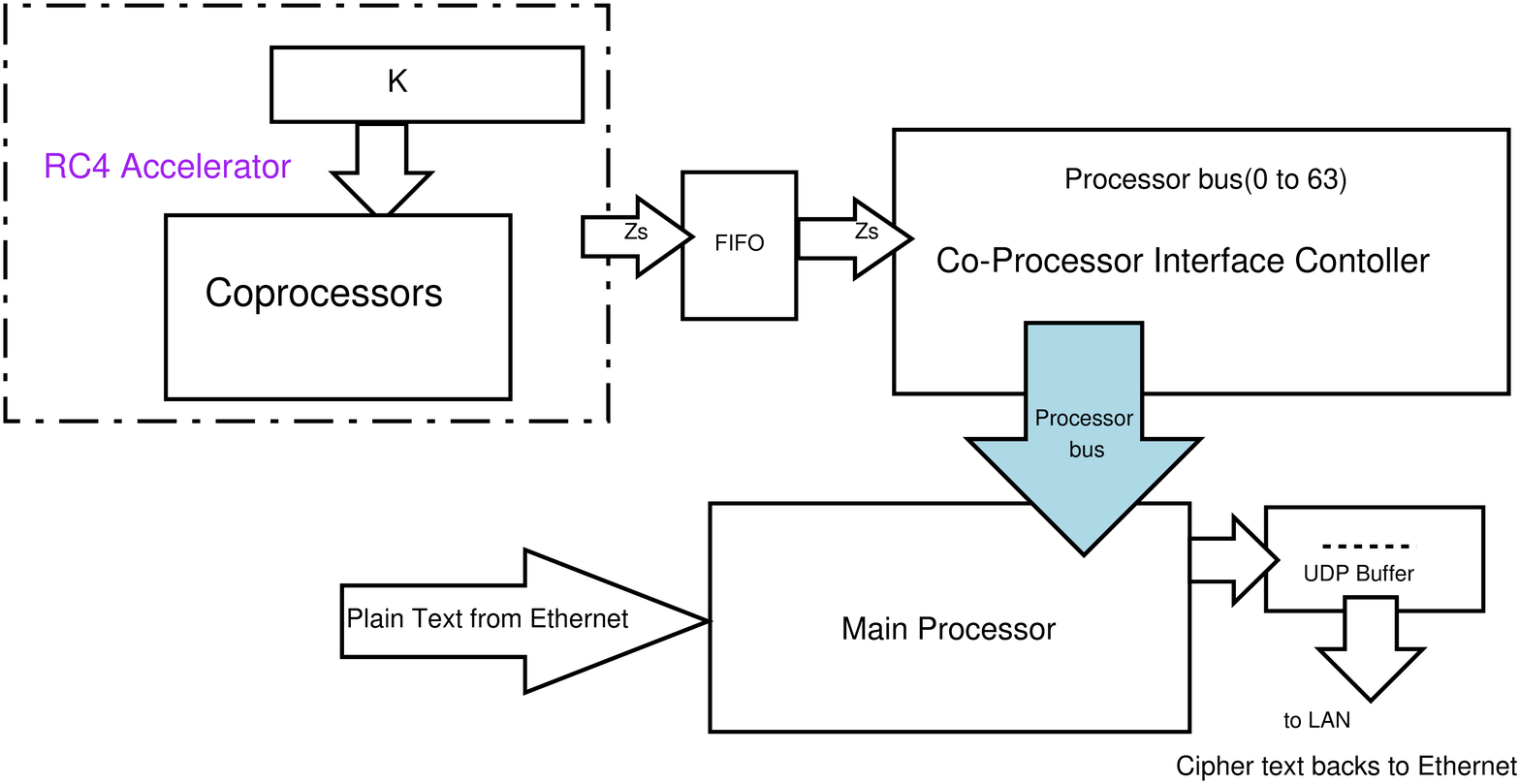}
\vspace{-6pt}
\caption{2 RC4 co-processor with main processor}
\vspace{-10pt}
\label{fig:4_co}
\end{figure}

The coprocessor and main processor work in truly parallel fashion which enhances the overall performance of the system. When plain text comes form RS232 to main processor, it sends a  Key request to coprocessor. The coprocessor generates Zs and stores into a FIFO. The main processor extracts the Zs from FIFO and xored with plan text. At the last phase the cipher text is sent to UDP buffer for the further ethernet processing as shown in Fig. \ref{fig:4_co}.
The results of consumption of hardware resources and electrical power of all the designs are noted at the simulation level and are presented in Sec. \ref{4.1}.Results of Throughput and a Comparative Study of it with that obtained with the existing implementations are presented in Sec. \ref{4.2}.
\vspace{-10pt}
\subsection{Results of Consumption of Hardware Resources and Power}
\label{4.1}
In this paper systematically accelerating 7 design algorithms of RC4 are proposed, such as, (D1) 1-byte per clock with 1 S-Box, (D2) 1-byte per clock with CKP and 1 S-Box, (D3) 2-bytes per clock with CKP and 1 S-Box, (D4) 2-bytes per clock with CKP and 2 S-Boxes, (D5) 4-bytes per clock with CKP and 2 S-Boxes, (D6) 4-bytes per clock with CKP and 4 S-Boxes, (D7) 8-bytes per clock with CKP and 4 S-Boxes.  All the designs are ported on Virtex 5 FPGA where the main processor handling the data interfaces functions in parallel with the co-processors undertaking parallel data processing and both together take care of the execution of RC4 algorithm.  The power consumption and hardware usage of the 7 designs are shown in Tables 2 and 3 respectively.   
Comparing D2 with D1, both having throughput of 1-byte-in-1-clock, it is noticed that D2 with CKP consumes lesser static power and generates lesser dynamic power (vide Table 2) and consumes substantially lesser silicon slices and LUTs (vide Table 3). For this reason the CKP has been incorporated in all future designs. Comparing D3 with D4, both having throughput of 2-bytes-in-1-clock, it is observed that D3 consumes lesser static power per byte, but generates more dynamic power per byte (vide Table 2), while the hardware usage for D3 on different accounts are substantially lesser (vide table 3). For generating 2 bytes per clock, the consideration of power and hardware usage indicates that D3 is always preferred to D4 for embedded system. On comparing D5 and D6, both having throughput of 4 bytes per clock, identical conclusion would also be drawn in favour of D5 (vide Tables 2 and 3). Hence to achieve a suitable throughput in hardware, the implementation of 2-bytes-1-clock mode of design is always preferred to 1-byte-1-clock mode of design.
\begin{table}[!h]
\vspace{-10pt}
\caption{Power consumption of 7 designs} 
\vspace{-10pt}
\centering  
\resizebox{8cm}{!}{%
    \begin{tabular}{|c|c|c|c|c|c|c|   }
        \hline
~  & \multicolumn{3}{c|}{All powers in mw } & \multicolumn{3}{c|}{All powers in mw/byte}\\  
\cline{2-7} 
Designs&	Static          & Dynamic&	Total     &Static        &Dynamic        &Total\\
~&	 Power     &	Power                               &Power     &Power/byte& Power/byte &Power/byte\\\hline
D1&	970.8             &206.4                 &1177.2      &970.8    &  206.4  &1177.2\\\hline
D2&	904.87    	     &159.85                &1064.72   &	 904.87&159.85  &1064.72\\\hline
D3&	 930.01              &502.87                   &1432.88       &465& 251.44&716.44\\\hline
D4&	 959.88          &373.61                  &1333.48     &479.94&186.81 &666.74\\\hline
D5&	976.24           &	656.75               & 1632.99    &244.06&164.19  & 408.25\\\hline
D6&	 971.34          &	572.74               &1544.08    &242.84 & 143.19 &386.02\\\hline
D7&	1010.58          &	1226.70           &2237.28      & 126.32 &153.34  &279.66\\
\hline
    \end{tabular}
}
\vspace{-10pt}
\label{table:pw_table} 
\end{table}   
\par  Regarding power, one would notice from Table 2 that with increasing throughput the consumption of static power per byte decreases nonlinearly with increasing nonlinearity, while the generation of dynamic power per byte decreases nonlinearly with decreasing non-linearity exhibiting a tendency to asymptotically assume a fixed value at a higher throughput. Regarding hardware, one would notice from Table 3 that the slices and LUTs required for generation of 2-bytes per clock by the CKP coprocessor of D3, D5 and D7 are around 2136 and 15411 respectively, while the same required for generation of 2-bytes per clock by the other one coprocessor in D4 are around 2097 and 17194 respectively and the same required for generation of 2-bytes by each of the other three coprocessors in D7 are around 2097 and 17640 respectively.   The results are in reasonable conformity.   
\par The critical path of our proposed designs is reasonably good. For 1-byte-1-clock mode of designs D1, D2, D4 and D6 the critical path is observed as 4.127 ns indicating 242 MHz as the maximum usable clock frequency. The clock frequency used in all the implementations of 1-byte-1-clock mode of designs is 200 MHz.  For 2-bytes-1-clock mode of designs D3, D5 and D7, the critical path is observed as 5.15 ns indicating 194 MHz as the maximum usable clock frequency. The clock frequency used in all the implementations of 2-bytes-1-clock mode of designs is also 194 MHz. 
\vspace{-10pt}
\begin{table}[!h]
\caption{Hardware usage of 7 designs} 
\vspace{-10pt}
\centering  
\resizebox{8cm}{!}{%
    \begin{tabular}{|c|c|c|c|c|c|c|c|c|   }
       \hline
& \multicolumn{2}{c|}{For the} &\multicolumn{2}{c|}{1st  CKP}   & \multicolumn{2}{c|}{Other}&  \multicolumn{2}{c|}{Per other}\\
D     & \multicolumn{2}{c|}{Design} &  \multicolumn{2}{c|}{Coprocessor}&  \multicolumn{2}{c|}{coprocessor(s)}&  \multicolumn{2}{c|}{coprocessor}\\
\cline{2-9}
~&	Slices	&LUTs	&Slices	&LUTs	&Slices	&LUTs	&Slices	&LUTs\\
\cline{2-9}
~	&1&2&3&4&5&6&7&8\\\hline
~&(3+5)&(5+8)&&&(1-3)&(2-5)&&\\\hline
D1	&4173	&14588	&4173 &14588	&---	&---	&---	&---\\\hline
D2	&2094	&5430            &2094 &5430	  &---	&---	&---	&---\\\hline
D3	&2136	&15411	&2136	&15411	&---	&---	&---	&---\\\hline
D4	&4144	&32523	&2094 &5430	&2050	&27093	&2050	&27093\\\hline
D5	&4233	&32605	&2136	&15411&	2097	&17194	&2097	&17194\\\hline
D6	&8256	&63556	&2094	&5430	&6162	&58126	&2054	&19375\\\hline
D7	&8425	&68331	&2136	&15411	&6289	&52920	&2096	&17640\\
\hline
    \end{tabular}
}
\vspace{-10pt}
\label{table:hw_table} 
\end{table}  
%
%
\begin{figure}[!htb]
\centering
\vspace{-10pt}
\includegraphics[scale=0.45]{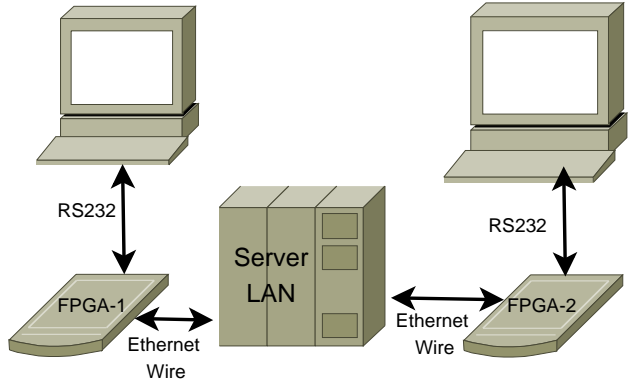}
\vspace{-10pt}
\caption{Experimental setup of FPGA based secured data communication}
\vspace{-10pt}
\label{fig:b_to_b}
\end{figure} 
\par All the designs has been implemented on Virtex 5 (ML505, lx110t) and Spartan 3E (XC3S500e) FPGA board using ISE 14.4 and EDK 14.4 tool. The power and resource result shown in Tables 2 and 3 have been obtained for the said Virtex board. The coprocessor has been coded by VHDL language and the main processor functionality has been projected by system C language.
Two Xilinx FPGA Spartan 3E (XC3S500e-FG320) boards, each with RC4 encryption and decryption engines separately, shown in Fig \ref{fig:b_to_b}, are connected through Ethernet ports for data transaction and each one to respective hyper terminals through RS-232 ports for display.
\vspace{-10pt}
\subsection{Throughputs of 7 Designs: Comparative Study with Existing Implementations}
\label{4.2}
In Table 4 the throughputs of all the existing works available in literature are given along with the same for all the 7 designs proposed for D1 through D7. From the Table 4, it is seen that the woks presented in refs. [21] and [20] are 3-bytes per clock designs, while those given in refs. [1] and [22], and also those presented in D1 (vide Sec. 3.1.2) and in D2 (vide Sec. 3.1.3), are essentially 1-byte per clock design. In D1 and D2, KSA process is executed within 257 clocks with an initial lag of 1 clock and the PRGA computes 1 byte per clock after a lag of 2 initial clocks. For D3 presented in Sec. 3.2.5, the KSA iterates 256 loops within 128 clocks and the PRGA executes 2-bytes in single clock after a lag of 2 initial clocks.  Although the PRGA throughput of designs D3 and D4 and that for ref [2] are identical, the number of clock to generate one byte for each of them are (1/2+642/n), (1/2+1284/n) and (1/2+259/n) respectively. This indicates that the design D3 is a better design than D4. The designs D5 and D6 have a throughput of 4 bytes per clock and the design D7, 8 bytes per cycle.  The additional PKRS process needs 1024 clock cycles for designs D2, D4 and D6 and 512 clock cycles for designs D3, D5 and D7. It is to be noted that the number of clocks needed to generate 1 byte for design D5, D6 and D7 are better than all existing RC4 architecture.  It has been seen that the two fastest RC4 with 1-byte and 2-bytes architectures get 14.8 Gbps and 30.72 Gbps throughput respectively in 65 nm ASIC platform where clock frequencies for 1 byte [1] and 2 byte architecture [2] are 1.85GHz and 1.92 GHz respectively. The designs D6 and D7, the fastest architecture proposed in our paper, get 5.96 Gbps and 11.57 Gbps throughput respectively in 65 nm FPGA technology where clock frequencies are 200 MHz and 194 MHz respectively.  Between the ASIC and the FPGA implementations, the commonalities are the 65 nm technology and the Gbps throughput, whereas the difference is between GHz and MHz clock frequencies.  The ASIC based designs, in comparison to FPGA based designs, achieved about 2.5 times higher throughput using about 9 times higher clock frequencies. The observation indicates that our proposed FPGA based parallel processing design algorithms to accelerate RC4 have some inherent strength. If the 2-bytes-1-clock mode of parallel processing design with 4 S-boxes (as in D7) generating 8-bytes per clock is implemented in ASIC 65 nm technology with 1.92 GHz of clock frequency, there is every likelihood to achieve substantially higher throughput.  
\begin{table}[!ht]
\vspace{-10pt}
\caption{Comparison of various performance metrics with existing designs} 
\vspace{-10pt}
\centering  
\resizebox{12cm}{!}{%
    \begin{tabular}{|c|c|c|c|c|c|c|}
        \hline
   Architecture   & \multicolumn{6}{c|}{Number of Clock Cycles } \\  
\cline{2-7}  
 ~                                             & KSA  &KSA          &   PRGA         &PRGA          &RC4 for & per byte o/p  \\ 
 ~                                             &(+PKRS*)       &per byte    &for n Bytes &for per bytes   & n byte & from RC4 \\\hline  
              
Ref  \cite{IEEE:b}, \cite{patent:matthews}&256X3=768  &3 & 3n & 3                    &3n + 768& 3+$\frac {768}{n}$ \\\hline
Ref  \cite{dp:math}	                &3+256=259	&1 +$\frac{3}{256}$   &3+n          &1+$\frac{3}{n}$       &259+(3+n)&1+$\frac{262}{n}$ \\\hline

Ref. \cite{springerlink:one_byte}       & 1+256=257	            &1 + $\frac{1}{256}$  &	 2+n                        & 1 + $\frac{2}{n}$                         &257+(2+n)                   &1+$\frac{259}{n}$ \\\hline
Ref \cite{ieee:two_byte}                       & 1+256=257                   &1+$\frac{1}{256}$    &$\frac{n}{2}$+2        &$\frac{1}{2}$+$\frac{2}{n}$         &257+(2+$\frac{n}{2}$)  & $\frac{1}{2}$+$\frac{259}{n}$ \\\hline
Design 1 	                             & 1+256=257	            &1+$\frac{1}{256}$    &2+n                         &1+$\frac{2}{n}$                           &257+2+n                      & 1+$\frac{259}{n}$    \\ \hline
Design 2*	                             & 1+256+	            &5+$\frac{2}{256}$    &2+n                         &1+$\frac{2}{n}$                           &1282+2+n                      & 1+$\frac{1284}{n}$           \\
~	                             & 1+1024=1282            &~    &~                         &~                        &~                      & ~           \\ \hline
Design 3*	                                     &1+128+	                   &2+$\frac{130}{256}$ &$\frac{n}{2}$+2  &$\frac{1}{2}$+$\frac{2}{n}$ &642+2+$\frac{n}{2}$      &$\frac{1}{2}$ +$\frac{642}{n}$            \\
~	                             & 1+512=642          &~    &~                         &~                        &~                      & ~          \\ \hline
Design 4*	           &1+256+                  &5+ $\frac{2}{256}$  &$\frac{n}{2}$+2        &$\frac{1}{2}$+$\frac{2}{n}$          &1282+2+$\frac{n}{2}$   & $\frac{1}{2}$ +$\frac{1284}{n}$    \\
~	                             & 1+1024=1282          &~    &~                         &~                        &~                      & ~           \\ \hline
Design 5*	           &1+128+	                   &2 + $\frac{130}{256}$ 	                 &~$\frac{n}{4}$+2     &$\frac{1}{4}$+$\frac{2}{n}$          &642+2+$\frac{n}{4}$ & $\frac{1}{4}$ +$\frac{644}{n}$               \\
~	                             & 1+512=642          &~    &~                         &~                        &~                      & ~           \\ \hline
Design 6*	           &1+256+             & 5+ $\frac{2}{256}$ 	      &$\frac{n}{4}$+2    &$\frac{1}{4}$+$\frac{2}{n}$          &1282+2+$\frac{n}{4}$ &$\frac{1}{4}$ +$\frac{1284}{n}$                \\
~	                             & 1+1024=1282          &~    &~                         &~                        &~                      & ~           \\ \hline
Design 7*	           &1+128+	                   &2+ $\frac{130}{256}$  &$\frac{n}{8}$+2          &$\frac{1}{8}$+$\frac{2}{n}$      &642+2+$\frac{n}{8}$ &   $\frac{1}{8}$ +$\frac{642}{n}$            \\
~	                             & 1+512=642          &~    &~                         &~                        &~                      & ~           \\ \hline

    \end{tabular}
}
\vspace{-30pt}
\label{table:compare} 
\end{table}
\vspace{-10pt}
\section{Study of Statistical Tests of long sequence of PRGA bytes of all Designs}
\vspace{-10pt}
\label{nist}
Of the 7 designs, from D1 to D7,  being studied on accelerating RC4, six, from D1 to D6, are studied at the implemented level following the simulation level studies, while D7 is being studied only at the simulation level since there has been a paucity of bus width for its implementation in FPGA. D1 is a conventional implementation of RC4 in 1-byte-1-clock mode of data processing without CKP, D2 is the same implementation with CKP and D3 doubles the throughput by processing 2-bytes in 1-clock preserving the CKP part of D2 intact. It may be noted that all the designs have identical 16-character key, while each of all the first three designs, D1, D2 and D3, has one S-Box, each of the next two designs, D4 and D5 has two S-boxes and the last two, D6 and D7, has four S-boxes.  Moreover, D1, D2, D4 and D6 are 1-byte-1-clock design, while D3, D5 and D6 are 2-bytes-1-clock designs.  For D1, the S-box that starts generating PRGA bytes is the post-KSA S-box, while the same for D2 and D3 are the post-PKRS ones.  Hence, the random key sequences for D2 and D3 are expected to be identical, while the same for D1 would be different. The design D4 uses two parallel S-boxes and generates 2 bytes in 1-clock, each of which is simultaneously contributed by each of the two S-boxes and is being processed at the output sequentially. In the next clock cycle another 2-bytes are generated in an identical manner giving rise to a sequence of 4-bytes in 2 clocks of which 1st and 3rd are from the 1st S-box, while $2^{nd}$ and 4th are from the $2^{nd}$ S-box. In D5, 4-bytes are generated in 1-clock of which the first two are simultaneously extracted from the two S-boxes during the first-half-clock of a clock cycle and the other two are also drawn simultaneously from the same two S-boxes during its second-half.  In nutshell, D4 generates a sequence of 4 bytes in 2 clocks, while D5 generates the same sequence in 1 clock. The same is true for D6 and D7 where both are using 4 S-boxes in which first set of 4-bytes are obtained simultaneously from the four S-boxes while the second set of 4 bytes are also obtained simultaneously from the same four S-boxes. It is therefore observed that the random key sequences are different to each other for D1, D2, D4 and D6 designs, while that for D3, D5 and D7 are identical to D2, D4 and D6 respectively.  Hence, statistical tests are to be undertaken for the four sets of random key sequences obtained from D1, D2, D4 and D6 designs in order to understand their randomness properties. Here for statistical tests, the indigenously coded NIST Statistical Test Suite comprising of 15 statistical tests \cite{nist:last}, \cite{sadique:nist} are adopted. 
\vspace{-10pt}  
\subsection{The Strategy of NIST Statistical Test Suite}
\vspace{-10pt}
The NIST Statistical Test Suite has 15 tests which are given in Sec. 5.2. The principal strategy of the NIST Statistical Test Suite is to judge statistical randomness property of random bit generating algorithms.  Based on 300 to 500 different keys, the algorithm generates a series of even number of different long random sequences of n bits, n varying between 13 and 15 lacs, each of which is tested by the 15 tests. Each test has a specific statistic parameter for bit sequences under the assumption of randomness and calculates the deviation of the respective statistic parameter for the series of tested bit sequences.   
\par From broad theoretical consideration, the 15 tests can be categorized into four groups; the first includes tests 1 to 4 as Frequency Tests, the second, tests 5 and 6 as  Repetitive Patterns Tests, the third, tests 7 to 12 as Patterns Matching Tests and the 
fourth, tests 13 to 15 as Random Walk Tests. 
\par It is noted that 6 tests (1, 3, 6, 9, 13 and 15) consider the entire bit sequence together and the theoretical studies available in literatures mention their statistic parameters and also the ways to calculate their respective simple deviations which is used in Error Function for 5 tests (1, 3, 6, 9, and 15) and in Normal Gaussian Distribution Function for test 13 to compute respective P-values.  
Other 9 tests (2, 4, 5, 7, 8, 10, 11, 12 and 14) consider the issue of degrees of freedom and the available theoretical studies indicate their respective statistic parameters and also mention ways to calculate their chi-square ($\chi^2$) deviations with degrees of freedom which are used in Gamma Function to compute their respective P-values. The 2 tests (2 and 7) consider the entire bit sequence being divided in N blocks and their P-values are computed based on Gamma Function with degrees of freedom N. The 4 tests (4, 5, 8 and 10) consider the entire bit sequence being divided in N blocks and also consider (K+1) classes obtained from respective theoretical studies and their respective P-values are computed using Gamma Function with degrees of freedom K, instead of N. The 3 tests (11, 12 and 14) consider the entire bit sequence together and introduce (K+1) classes obtained from respective theoretical studies while calculating their statistic parameters which are used in Gamma Function with degrees of freedom K in order to compute their P-values. The methodologies to compute P-values using Error function, Normal Gaussian Distribution function and Gamma function are well described in \cite{nist:last}, \cite{sadique:nist}.
\vspace{-10pt}
\subsection{Brief description of test-wise strategy to evaluate deviation of statistic parameters}
\vspace{-10pt}
The purposes in each test being applied on n-bit binary sequence are briefly narrated focusing the statistic parameters and methods to calculate the simple deviations or the  $\chi^2$-deviations as the case may be.\\
\textbf{Test 1:  Mono-bit Frequency test:} It is intended to see if the difference of number of 0s and 1s falls within the limit of randomness. From $\varepsilon_i$ binary sequence, different $\chi_i$ sequence using a mathematical relation $\chi_i$ = 2$\varepsilon_i$ - 1 is obtained where 0s are changed to -1 keeping 1s as +1. The summation of |$\chi_i$| gives the simple deviation of magnitude of the difference of number of 1s and 0s is calculated. The P-value is computed by using the simple deviation in Error function.\\
\textbf{Test 2: Mono-bit Frequency test in N blocks:} The purpose is to see if the number of 1s in M-bit N Blocks is close to M/2 and the $\chi_2$-deviation of proportion of 1s in M-bit blocks from probability 1/2 is calculated. The P-value is computed by using the $\chi^2$-deviation in Gamma function with N degrees of freedom.\\ 
\textbf{Test 3: Test of Runs of 0s and 1s:} It is intended to see if the frequencies of runs of 1s and 0s of various lengths across the entire sequence are in limits of randomness and the simple deviation of number of runs of 0s and 1s together from that obtained for random sequence is calculated. The P-value is computed by using the simple deviation in error function.\\
\textbf{Test 4: Test of Longest runs of 1s in N blocks involving (K+1) classes:} It is intended to see if the frequencies of (K+1) classes of longest runs of 1s appearing in the N-block sequence each of M bits falls within the limit of randomness.  From relevant theoretical studies it is observed that K is considered as 3 for M=8 and there are 4 class of blocks ($\nu_i$, $i$ = 0 to 3) based on longest run of 1s varying from 1-bit or less to 4-bit or above and the $\chi^2$-deviation of the observed four classes from expected ones can be calculated for M=8.  The theoretical studies also indicate that 6 classes can also be considered for larger values of M, e.g. 128, 512 and 1000 and 7 classes if M = 10000.   The P-value is computed by using $\chi^2$-deviation in Gamma function with K degrees of freedom.              \\       
\textbf{Test 5: Binary Matrix Rank test in N blocks involving 3 classes:} The test intends to see linearly dependent repetitive patterns within its fixed length N blocks of sub-strings of M2 bits forming (M x M) matrix and to search its rank. The choices of M ($\le$32) and N are made in such a way that the discarded bits (n - $M^2$.N) become least. The search is for the rank of an M-order matrix by calculating determinants of all its possible sub-matrices including the own one. If the determinant of M-order matrix is non-zero, its rank is M; if not and if at least the determinant of one of (M-1)-order sub-matrices is non-zero, its rank is (M-1); if not, one has to look for all possible sub-matrices of (M-2)-order and downwards. For a full rank sub-matrix, there are no repetitive patterns. From relevant theoretical studies under the assumptions of randomness for M $\ge$ 10, the probability data of full rank M-order matrix and other lower order sub-matrices are available. The probability values for sub-matrices of (M-2)-order and lower have been observed to be so low that these can be clubbed together without any effective loss of information providing three classes of matrices and giving 2 degrees of freedom from class point of view for the distribution function. The $\chi^2$-deviation of full rank and rank-deficiency matrices from theoretically expected ones gives deviation values of the statistic parameter. The P-value is computed by using the $\chi^2$-deviation in Gamma function with 2 degrees of freedom. \\
\textbf{Test 6: Discrete Fourier Transform test:} The purpose of the test is to detect periodic features in the n-bit sequence by focusing on the peak heights in its Discrete Fourier Transform.  The periodic features indicate the presence of repetitive patterns that are close to each other. It is intended to see the test statistic parameter (d), which is the normalized difference between the observed numbers of peaks exceeding 95\% threshold (T) and the expected theoretical number of such peaks under the assumption of randomness, is within the limit of randomness when less than 95\% peak occur below T and greater than 5\%, above T.  The P-value is computed by using d in Error function.         \\       
\textbf{Test 7: Non-overlapping Template test in N Blocks:} The purpose of the test is to search number of occurrences of pre-defined non-periodic (aperiodic) patterns by using a non-overlapping m-bit window sliding by 1-bit - once the pattern is found, the window is reset after sliding m-bit.  The entire n-bit sequence is divided in N blocks each of M bits where M is expected to be greater than 1\% of n and N is the integer division of n by M. The distribution of such non-periodic patterns in blocks is theoretically found to follow Gaussian distribution function based on Central Limit theorem and expressions of mean and standard deviation of number of occurrences of all specified pattern are also given following exhaustive theoretical studies.  The number of times the specified patterns do occur in each block is counted and the $\chi^2$-deviation of the same from the mean is obtained. The P-value is computed by using the $\chi^2$-deviation in Gamma function with N degrees of freedom. \\
\textbf{Test 8: Overlapping Template test in N Blocks involving 6 classes:} The purpose of the test is to search number of occurrences of pre-defined non-periodic patterns by using an overlapping m-bit window sliding always by 1-bit irrespective of whether the pattern is found or not and to see if the number of such occurrences is within the limit of randomness obtained from relevant theoretical studies. The sequence is divided in N blocks each of M-bit in such a way that N is the integer division of n by M and is at least greater than 5.  Six classes $\nu_i$ with 0$\le$i$\le$5, is defined as the number blocks where m-bit pattern is found in i-times. It is found that for this test the Poisson asymptotic distribution is being followed. From theoretical studies the probability ($p_i$) of occurrences of $\nu_i$ blocks corresponding to 6 classes are available in literature.  The statistic parameter is the observed $\nu_i$ blocks for 6 classes and its  $\chi^2$-deviation is calculated based on observed and expected proportion of $\nu_i$ blocks for 0 $\le$ $i$ $\le$5.  The P-value is computed by using the $\chi^2$-deviation in Gamma function considering 5 degrees of freedom.  \\     
\textbf{Test 9: Maurer's "Universal Statistical" test:} The test checks if the sequence is a significantly compressible one without any loss of information. A significantly compressible sequence is considered to be non-random. The focus of the test is to measure the distances in number of bits between two locations of an L-bit pattern, one in the initialization segment Q and others in the test segment K. Such distances are measures that are related to the lengths of compressed sequence. Out of the n-bit sequence, the initialization segment has (QL) bits and the test segment has (KL) bits discarding less than L bits, L can take values between 6 and 16.  The test statistic parameter fn(L) is the average of all such distances in number of bits arising out of the presence of the L-bit pattern in many places of K test segments. Under the assumption of randomness, the Standard Normal Gaussian Distribution function is assumed and an expression of expected value of the test statistic parameter$ Ef_n(L)$ is available in literatures  including an expression of variance(L) which also finds the standard deviation ($\sigma$) as a function of K. The simple deviation is calculated based on fn, $Ef_n(L)$ and $\sigma$. The P-value is computed by using the simple deviation in Error function. \\  
\textbf{Test 10: Linear Complexity test in Blocks involving 7 classes:} The purpose of the test is to see if the sequence is complex enough to be considered random.  The focus of the test is the length of a Linear Feedback Shift Register (LFSR) which generates the sequence.  The bit sequence from which a longer LFSR is obtained can be termed as random, while the shorter LFSR indicates non-randomness. The Berlekamp-Massey Algorithm is adopted to obtain a LFSR. The n-bit sequence is divided into N blocks, each of M bits.  Under the assumption of randomness, the theoretical mean length of LFSR ($\mu$) and also the statistical deviation, Ti of a LFSR of length Li are obtained from literature and seven range of values of Ti are marked between -2.5 and +2.5 as 7 classes for 0 $\le$ $i$ $\le$ 6. The theoretical probability ($\pi$) of number of blocks ($\nu_i$) corresponding to each of the 7 classes is also obtained from standard literature.  Following the extraction of LFSR of each M-bit N blocks and its length (Li), the test statistic parameter is marked as the number of blocks ($\nu_i$) for 0 $\le$ $i$ $\le$ 6 and the $\chi^2$-deviation is calculated followed by computation of P-value by using the  $\chi^2$-deviation in Gamma function and considering 6 degrees of freedom.            
\textbf{Test 11: Serial test involving classes:} The test introduces a battery of procedures and intends to see the uniformity of distributions of patterns of some given lengths. The purpose of the test is to determine if number of occurrences of $2^m$ m-bit overlapping patterns is approximately the same, as would be expected for a random sequence. The values of m is small, less than or equal to [int($log_2$n)-1]. Under the assumption of randomness, every m-bit pattern has the same chance of appearing as every other m-bit patterns and each of all m-bit patterns is expected to occur Am times, where $A_m$ = $n/2^m$. Across the entire bit sequence, the frequencies $\nu_i$(m) of all possible overlapping m-bit patterns are counted for all decimal values of $i$ from $i$=0 to $i$ $\le$ $(2^m-1)$ and $\psi_m^2$-deviation of $\nu_i(m)$ from $A_m$ is calculated. Similarly, $\psi_{m-1}^2$-deviation and $\psi_{m-2}^2$-deviation are also calculated for (m-1)-bit and (m-2)-bit patterns considering $A_{m-1}$ = n/$2^{m-1 }$ and $A_{m-2}$ = n/$2_{m-2}$ followed by calculating 1st order difference of $\psi_m^2$, i.e. $\Delta \psi_m^2$ = $\psi_m^2$ - $\psi_{m-1}^2$ and 2nd order difference of $\psi_m^2$, i.e. $\Delta^2\psi_m^2$ = $\psi_m^2$ - 2$\psi_{m-1}^2$ + $\psi_{m-2}^2$. The $\Delta$$\psi_m^2$ has a $\chi^2$ distribution with $2^{m-1}$ degrees of freedom and $\Delta^2$$\psi_m^2$ has also a $\chi^2$ distribution with $2^{m-2}$ degrees of freedom and the Serial test produces two P-values using Gamma function. For m=1 the Serial test turns out to be the Frequency Test. \\  
\textbf{Test 12: Approximate Entropy test using m-bit Pattern involving $2^m$ possibilities:} Entropy is a feature of randomness of an-bit sequence involving repetitive patterns. To measure entropy ($\phi^m$) of a sequence involving all possible overlapping m-bit patterns across the entire sequence, (m-1) bits are copied from the beginning and appended at the end of the sequence and the relative frequencies, $\pi_i$(m) are calculated considering $i$ as the decimal values of all possible m-bit patterns over (n+m-1) bit sequence followed by calculation of $\phi^m$ as the summation of [$\pi_i$(m)ln  $\pi_i$(m)] for $i$ varying from 0 to $2^m$-1. Similarly, $\phi^{m+1}$ is also calculated using a (m+1)-bit patterns for the same sequence over (n+m) bit sequence. The values of m and n are to be so chosen that m becomes less than [int$(log_2n$)- 5]. The test statistic parameter is the difference of $\phi^m$ and  $\phi^{m+1}$ which is also termed as approximate entropy \textit{$A_pE_n(m)$}. For a random sequence, the  \textit{$A_pE_n(m)$} assumes a maximum value projected as ln2. The smaller values of \textit{$A_pE_n(m)$} indicate strong regularity implying non-randomness, while its larger values indicate substantial fluctuation or irregularity implying randomness.  The$\chi^2$-deviation of the statistic parameter is calculated by comparing its value from ln2.  The P-value is computed using $\chi^2$-deviation in Gamma function with $2^m$ degrees of freedom. \\
\textbf{Test 13: Cumulative Sums (cusum) test:}  This test intends to see if either 1s or 0s appear in large numbers at early stages and at later stages of the tested sequence and for these two partial sequences of about 100 bits or more are considered at the beginning and at the end respectively. The expression X = 2$\epsilon_i$ - 1, used in Test 1, makes the \{$0,1\}^n$$\epsilon_i$ sequence to \{-1,+1\}$^n$ $\chi_i$  sequence. The cumulative sums of the adjusted \{-1,+1\} $X_i$ digits of a partial sequence at the beginning or at the end of the sequence is obtained as $S_i$ = $S_{i-1}$+ $X_i$ with $S_0$ = 0 and while on the run its maximum value z = max(|$S_i$|) is picked up. The purpose of the test is to see if the cusum of the partial sequences occurring in the tested sequence is too large or too small relative to the expected behavior of the same for random sequences. In the event the statistic parameter (z/$\sqrt{n}$) is large, the bit sequence is considered to be non-random.  It may be noted the cusum can be undertaken in a forward manner (mode=0) starting from the beginning of the sequence to the end of the partial sequence and a larger statistic value indicates that 1s or 0s occur in large numbers at the early stages of the sequence. The cusum can also be undertaken in a reverse manner (mode=1) starting from the end of the sequence to the beginning of the partial sequence and if its statistic value is large, 1s or 0s do occur in large numbers at the end of the sequence.  Small values of the statistic for both the modes indicate that 1s and 0s are intermixed evenly across the entire sequence. For each of the two modes mentioned above, two z values are noted. It is observed that the distribution of cusum follows the Standard Normal Gaussian Distribution function following which the two P-values are computed.  \\
\textbf{Test 14: Random Excursions test involving 6 visits:} The test intends to ensure that 1s and 0s do occur intermingled evenly across the entire sequence including its mid-region as it happens in random sequences. 
As is being done in tests 1 and 13, the $\{0,1\}^n$ binary sequence ($\epsilon_i$) is transformed to $\{-1, +1\}^n$ sequence ($X_i$ ). In the preset test a cusum state sequence $S_i$ is obtained using the expression $S_i$ = $S_i$-1+ $X_i$ with $S_0$ = 0 for $i$ = 1 to n. The sequence $S_i$ is now bounded by two zeros at the beginning  and at the end.  The states, e.g. $\pm$1, $\pm$2, $\pm$3, $\pm$4, $\pm$5, $\pm$6 and $\pm$7 including other intermediate zeros appearing on the bounded sequence. If the random excursion path of the bounded Si is graphically traced on the X-Y plane from (0,0) to (0,n+1), one observes that the trace starts from the origin (0,$S_0$) and ends at (0,$S_{n+1}$) and goes through many intermediate zeros where it has intercepted the X-axis indicated the 0-states of $S_i$. Between two successive 0-states there is a cycle, called a J-cycle and across the entire sequence there are many J-cycles involving various positive and negative states. 
The issue that is considered in the test is the number of J-cycles in which, say, the appearance for state +1 takes place for k-times in $\nu_k$(+1)-cycles for 0$\le$k$\le$5, k=0 means no appearance and $\sum$$\nu_k$(+1)=J(+1)-cycles. Similarly one can find all the eight J($\pm$s)-cycles. The theoretical studies define $\pi_k(s)$ as probability of J-cycles k-times visits to a state s and provide separate algebraic expressions for k=0, k=1 to 4 and k > 5. The theoretical results indicate that in 90\% and above cases states 5, 6 and 7 are found in no J-cycles.  This is the reason that theoretical probabilities $\pi_k(s)$ are considered for k varying from 0 to 5 with 5 degrees of freedom.  In the event the states $\pm$6 and $\pm$7 are found k-time in J-cycles, these are included $\nu_5(s)$ for appropriate values s. Hence from practical consideration the eight states, e.g. $\pm$1, $\pm$2, $\pm$3 and $\pm$4 are considered and the eight $\chi^2$-deviation considering differences between corresponding to $\nu_k$ (observed) and $\pi_k$ (expected) are calculated and using them in Gamma function eight P-values are computed considering 5 degrees of freedom. If J < max(0.005$\sqrt{n}$, 500), the sequence is considered to be non-random. A one million bit sequence is considered non-random if J is less than 500.  Among eight P-values, if failing P-values are very much in minority, i.e. 1 or 2, the test can be considered to pass. \\
\textbf{Test 15: Random Excursions Variant Test:} The test looks for number of visits, $\varepsilon(s)$ to a particular state, s in cusum random walk across the entire bit sequence and estimates deviations from expected number of visits in the random walk under the assumption of randomness. The 2nd paragraph of Test 14 is also applicable in Test 15.  The statistic parameter in Test 14 is the number of J-cycles involved in making visits to a state for k-times where k varies from 0 to 5, while the same in Test 15 is the number of visits, $\varepsilon(s)$  to a particular state across the entire bit sequence involving J-cycles in totality.  Simple deviation of  $\varepsilon(s)$  from J is calculated under the assumption that  $\varepsilon(s)$  follows the Standard Normal Distribution function whose Statistic Variation $\sigma$ is given as (4|s|-2) for a particular state s.  The simple deviation is used in Error function and P-value is computed for a particular state visit s. There are 18 states, e.g.  $\pm$9,  $\pm$8,  $\pm$7,  $\pm$6,  $\pm$5,  $\pm$4,  $\pm$3,  $\pm$2,  $\pm$1 and the observation on the 18 states would give us 18 P-values. 
\vspace{-10pt}
\subsection{ Statistical aspect of P-values in NIST Statistical Test Suite}
\vspace{-10pt}
Based on computed P-values, if one intends to draw statistically sound conclusion regarding randomness property associated with a random number generating algorithm, one has to consider a large number of long binary sequences numbering between 300 and 1000, each one to be comprised of about 13 to 15 Lacs of bits.  It may be noted that of the 15 NIST tests, 11 tests (nos. 1, 2, 3, 4, 5, 6, 7, 8, 9, 10 and 12) give 1 P-value each, each of the 2 tests(nos. 11 and 13) give 2 P-values, test 14, 8 P-values and test 15, 18 P-values. According to NIST Test Suite, the individual passing criterion of a P-value is to be set, the statistical criterion-1is to be set to pass majority of P-values obtained by a test applied on all data files generated by an algorithm and then the statistical criterion-2 is to be set to judge the uniformity of distribution of all the P-values between 0 and 1 considered previously. The P-values (pv) are grouped in 11 ranges as,
 ($C_0) 0 \leq pv < 0.01,~~(C_1) 0.01 \leq pv < 0.1,~~(C_2) 0.1 \leq pv < 0.2,~~ (C_3) 0.2 \leq pv < 0.3,~~(C_4) 0.3 \leq pv < 0.4,~~(C_5) 0.4 \leq pv < 0.5,~~(C_6) 0.5 \leq pv < 0.6,~~ (C_7) 0.6 \leq pv < 0.7,~~ (C_8) 0.7 \leq pv < 0.8,~~ (C_9) 0.8 \leq pv < 0.9$ and $(C_{10}) 0.9 \leq pv < 1.0$. 
and the numbers in each group are noted. It is statistically reasonable for some P-values to fail. If an algorithm always generates data files giving rise to failing P-values, the algorithm always generates non-random numbers. The same is also true if an algorithms generates passing P-values clubbed in a specific range. The non-uniformity in the distribution of P-values is a non-random feature.  \\
The three statistical criteria said above can be expressed as follows:
\textbf{Individual Passing Criterion of P-value (Significance level = 0.01):} A P-value is considered to pass if it is greater than $\alpha$. A datafile can be considered to pass a test having one P-value, if its P-value is greater than $\alpha$.The same data file can also be considered to pass another test having multiple P-values, if all its P-values are greater than $\alpha$. 
\textbf{Statistical passing criterion:} Statistical passing criterion of a test applied on f number of data files generated by an algorithm: The statistical passing criterion is expressed by a threshold parameter, named as Expected Proportion Of Passing, abbreviated as EPOP. For a test that considers f data files, the EPOP is calculated following the formula given below:
\vspace{-10pt}
\begin{equation}
\label{epop}
EPOP=(1- \alpha)-3 \times\sqrt{\frac{\alpha(1-\alpha)}{p}}
\vspace{-10pt}
\end{equation}
       where p = total no. of P-values for the concerned test = no. of P-values evaluated for one datafile multiplied by f. After applying the criterion-1 on all the data files by a test and storing the number of P-values falling within the range of 11 columns stated above, one can estimate Observed Proportion Of Passing (OPOP) as the ratio of number of passing P-values to the total number of P-values and can compare the same with EPOP.The concerned test passes the statistical passing criterion, if OPOP > EPOP. 
\textbf{Statistical Uniformity Criterion:} Statistical Uniformity Criterion of P-values generated by a test applied on data files generated by an algorithm:The statistical uniformity criterion of all the P-values by a test is expressed by a parameter, called as P-value of P-values and abbreviated as POP. The range of P-values from 0 to 1 is considered being divided in 10 columns from $C_1$ to $C_{10}$ as stated above and the number in $C_0$ is considered being added with C1. Ideally the p number of p-values is uniformly distributed if each of the 10 groups has(p/10) number of P-values.  In a practical situation one has to estimate the chi-square $\chi^2$ deviation using the formula given below:
\vspace{-10pt}
\begin{equation}
\label{chi}
\chi^2=\sum\limits_{i=1}^{10} \frac{(S_i-\frac{m}{10})^2}{\frac{m}{10}}
\vspace{-10pt}
\end{equation}
The consideration of 10 ranges indicates the degrees of freedom as K = 9 and the POP can be calculated from the following formula involving Gamma function (a, x) where a = K/2 and x=$\chi^2$/2,
\vspace{-10pt}
\begin{equation}
\label{pop}
POP=1-\frac{\Gamma(a,x)}{\Gamma(a,\infty)}
\vspace{-10pt}
\end{equation}
If POP > 1e-4, the distribution of P-values is uniform for the concerned test. 
\vspace{-10pt}
\subsection{Results of Statistical Experiments} 
For undertaking Statistical experiments, 300 data files each of length 13,42,400 bits are generated by using 300 different 16-character keys. It is observed that for a particular key the generated data files for three pair of designs, namely D2-D3, D4-D5 and D6-D7 are identical. Using indigenously coded NIST Statistical Test Suite, statistical testings are undertaken on four sets of data files generated by the seven designs, namely D1, D2-D3, D4-D5 and D6-D7 and all the computed P-values are noted in a directory "Design-wise and Test-wise of P-values of 4 Designs" [link].
For histogram analysis of test-wise P-values of each of 4 designs, 11 columns as stipulated in Sec. 10.2 above are created and while reading the test-wise respective data from the directory said above, the data in the appropriate range of the 11 columns are noted in Table \ref{table:nist1} for the designs D1 and D2-D3, while those for the designs D4-D5 and D6-D7, in\ref{table:nist2}.  The test-wise EPOP data for all the 15 tests are calculated using eq.\ref{epop} and the same are noted in the 2nd column of Tables \ref{table:nist1} and \ref{table:nist2} for all the 5 designs.The test-wise OPOP data is calculated as the ratio of summation of the histogram data of P-values in 10 columns from C1 to C10 to summation of the same in the 11 columns from $C_0$ to $C_{10}$ and the ratios for all the 15 tests corresponding to  each of the 4 designs are noted in the 13th column of Tables\ref{table:nist1} and \ref{table:nist2}.  The 15th column notes the remarks of OPOP data as $R_O$ and each row of it marks the passing of the Statistical Passing criterion of a particular OPOP as 'Y', if it is greater than EPOP, else 'N'.  The marking of 'Y' means, a stipulated test considers a particular design algorithm to pass the statistical passing criterion.
Now one has to see if the histogram data of P-values obtained experimentally are uniformly distributed and satisfies the uniformity criterion.  The test-wise chi-square ($\chi^2$) deviation for a particular design is calculated using eq. \ref{chi} and the POP data (P-value Of P-values) is calculated using eq.\ref{pop}.The test-wise POP data for all the 15 tests corresponding to each of the 4 designs are noted in the 16th column of Tables\ref{table:nist1} and \ref{table:nist2}.  The 17th column notes the remarks of the POP data as $R_P$ and each row of it marks the uniformly passing of criterion of a particular POP data as 'Y', if POP is greater than 1e-4, else 'N'. The marking of 'Y' means that a particular test considers the concerned design algorithm to pass the statistical uniformity criterion. When both 2nd and 3rd the criteria  are satisfied for all the tests for a particular design, the concerned algorithm generates sequences of random numbers or bits that can be considered to pass the test of randomness and this indicates that the algorithm generates sequence of random numbers or bits with majority with passing P-values and all the P-values including the failing ones are distributed uniformly between 0 and 1. 
\vspace{-10pt}
\subsection{Discussion of Results of Statistical Tests}
\vspace{-10pt}
Looking at the $R_O$ and $R_P$ data marked respectively in 15th and 17th columns for D1 and D2-D3 in Table \ref{table:nist1} and D4-D5 and D6-D7 in Table \ref{table:nist2}, one observes 'Y' for all 15 tests of all the designs indicating that all the designs produce random sequences of bits including the conventional RC4 designed in D1 in 1-byte-1-clock mode. The S-box generated by the PKRS process for the design D3 in 2-bytes-1-clock mode increases its security without sacrificing the randomness feature of its generated 
bit sequences. For 1-byte-1-clock design D4 and 2-bytes-1-clock design D5, the PKRS process generates two S-boxes at the different stages of randomization of the same string of initial byte sequences each producing identical byte sequence with increased security and throughput with no loss of their randomization features.  The same is observed to be true for the 1-byte-1-clock design D6 and 2-bytes-1-clock design D7 for which the PKRS process generates four S-boxes at the different stages of randomization.                
\vspace{-20pt}
\section{Conclusion}
\vspace{-15pt}
\label{con}
The innovative design concepts in respect of adopting MUX-DEMUX based swaps, utilizing rising as well as falling edges of clocks followed by introducing suitable pipeline structures for optimum space-time management for data processing has led to achieve increased throughput. The PKRS process has contributed multiple S-boxes creating scopes to use inherent parallelization features available in FPGA based embedded systems for increased throughput with increased security and without sacrificing randomization features of its bit sequences.  The byte management coupled with clock management has been done in such a way that the sequence of bytes, being picked up sequentially from multiple S-boxes in designs with 1-byte-1-clock mode in 1-clock and in designs with 2-bytes-1-clock mode in 1/2-clock, become identical - the throughput is doubled without disturbing the byte sequence. The important thing is that the conventional RC4 is designed in hardware strictly by adopting 1-byte-1-clock mode of data processing at all stages and achieved a throughput of 1-byte in 1-clock  without adopting 2-bytes-1-clock mode of byte processing \cite{ieee:two_byte} in some stages. At present it has been possible to achieve a maximum possible throughput of 4-bytes in 1-clock while implementing designs D5 with 2 S-boxes and D6 with 4 S-boxes in Virtex 5 FPGA.  It may be noted that the silicon spaces necessary for implementing the logic circuits required for processing 2 bytes together in 2 S-boxes in D5 is much less that required for implementing 4 S-boxes and necessary 1-byte-1-clock circuits. With hardware limitation in existing FPGA it has been possible to implement 2-bytes-1-clock mode of design in D7 with 4 S-boxes in FPGA and to achieve a throughput of 8 bytes in 1-clock at the simulation level. It is foreseen, if a suitable FPGA board is available, it is possible to implement such a design of stream cipher using 8 S-boxes in embedded system capable to achieve substantial acceleration with increased security and increased throughput of 16 bytes in 1-clock. 
\begin{table}[!htbp]
\caption{Data for D1, D2 \& D3 } 
\vspace{-10pt}
\centering  
\resizebox{9cm}{!}{%

    \begin{tabular}{|c|c|c|c|c|c|c|c|c|c|c|c|c|c|c|c|c|}

\hline

T\# &       $C_0$&   $C_1$&   $C_2$&   $C_3$&   $C_4$&  $ C_5$&   $C_6$&  $ C_7$&   $C_8$&  $ C_9$&   $C_{10}$&   \small{EPOP}&      \small{OPOP}&   $R_{o}$& POP&    $R_{p}$\\ \hline
\multicolumn{17}{|c|}{D1}\\\hline

 01&     4&    27&    33&    28&    34&    28&    28&    30&    31&    25&    32&   0.973& 0.987  &	Y&   0.987 &Y\\\hline
 02&     2&    29&    34&    36&    26&    34&    35&    24&    26&    30&    24&   0.973& 0.993  &	Y&   0.679 &Y\\\hline
 03&     4&    23&    36&    33&    24&    33&    38&    29&    33&    28&    19&    0.973& 0.987  &	Y&  0.356 &Y\\\hline
 04&     3&    32&    28&    24&    36&    24&    32&    31&    32&    33&    25&    0.973 &0.99  &	Y&  0.74   &Y\\\hline
 05&     4&    29&    30&    17&    28&    36&    31&    33&    33&    37&    22&    0.973Y& 0.987  &	Y&  0.233  &\\\hline
 06&     5&    24&    43&    29&    28&    38&    19&    31&    29&    25&    29&    0.973& 0.983  &	Y&  0.166  &Y\\\hline
 07&     3&    36&    37&    22&    24&    28&    28&    34&    29&    29&    30&    0.973& 0.99  &	Y&  0.481  &Y\\\hline
 08&     3&    31&    34&    29&    36&    22&    21&    22&    39&    39&    24&   0.973 &  0.99  &	Y&   0.07   &Y\\\hline
 09&     3&    31&    29&    25&    22&    22&    32&    27&    40&    34&    35&    0.973&  0.99  &	Y&  0.29  &Y\\\hline
 10&     2&    23&    31&    32&    24&    20&    35&    32&    30&    31&    40&    0.973 &  0.993  &Y&  0.361  &Y\\\hline
 11&     8&    45&    63&    62&    65&    52&    72&    61&    54&    62&    56&   0.978 &  0.987  &Y&  0.753  &Y\\\hline
 12&     4&    21&    37&    34&    28&    25&    37&    30&    27&    30&    27&    0.973&  0.987  &	Y&  0.72  &Y \\\hline
 13&     2&    49&    60&    64&    64&    59&    65&    56&    77&    52&    52&    0.978 &  0.997  &	Y&  0.39  &Y\\\hline
 14&    27&   229&   215&   237&   232&   252&   226&   280&   238&   228&   236& 0.98&  0.989  &	Y4&  0.175  &Y\\\hline
 15&    50&   513&   579&   537&   573&   587&   550&   544&   504&   463&   500&  0.986&  0.99  &	Y&  0.002  &Y\\\hline

\multicolumn{17}{|c|}{D2, D3}\\\hline 
1&      1&    29&    28&    42&    24&    34&    28&    36&    24&    20&    34&   0.973& 0.997 &  Y& 0.16  &Y\\\hline
 2 &     3&    20&    31&    23&    28&    41&    42&    30&    24&    31&    27&   0.973&  0.99 &  Y& 0.13  &Y\\\hline
 3 &     6&    25&    23&    40&    29&    39&    31&    38&    20&    25&    24&   0.973&  0.98&  Y& 0.084  &Y\\\hline
 4  &    1&    37&    31&    29&    31&    31&    27&    32&    27&    26&    28&   0.973&  0.997&  Y& 0.932  &Y\\\hline
 5  &    2&    30&    22&    40&    30&    32&    24&    35&    30&    25&    30&  0.973&  0.993&  Y& 0.475  &Y\\\hline
 6 &     6&    35&    30&    30&    28&    40&    15&    39&    36&    19&    22&   0.973&  0.98&  Y& 0.003  &Y\\\hline
 7 &    5 &   28 &   34 &   36 &   26 &   30 &   25 &   29 &   29&    25  &  33 &  0.973&  0.983&  Y& 0.868  &Y\\\hline
 8 &     3&    28&    31&    34&    25&    33&    25&    38 &   31&    27 &   25&   0.973&  0.99&  Y& 0.753  &Y\\\hline
 9 &     2&    25&    32&    32&    36&    25&    32&    30 &   37&    26 &   23&   0.973&  0.993&  Y& 0.686  &Y\\\hline
 10 &    1&    30&    23&    27&    41&    25&    23&    28&    37&    37&    28&   0.973& 0.997&  Y& 0.213  &Y\\\hline
 11 &    4&    54&    58&    60&    60&    62&    60&    60&    66&    65&    51&  0.978 &  0.993&  Y& 0.979  &Y\\\hline
 12  &   2&    27&    34&    28&    29&    29&    28&    32&    30&    36&    25&   0.973&  0.993&  Y&  0.962  &Y\\\hline
 13 &   12&    63&    50&    49&    59&    62&    71&    63&    47&    64&    60&   0.978&  0.98&  Y & 0.173  &Y\\\hline
 14 &   18&   199&   238&   249&   260&   226&   232&   249&   246&   247 &  236& 0.984&  0.993&  Y& 0.733  &Y\\\hline
 15 &    69&   437&   503&   490&   559&   600&   559&   583&   565&   514&   521& 0.986&  0.987&  Y& 0.005  &Y

\\\hline
    \end{tabular}
}
\label{table:nist1} 
\end{table}

\vspace{-20pt}
\begin{table}[!htbp]
\caption{Data for D4, D5, D6 \& D7 } 
\vspace{-10pt}
\centering  
\resizebox{9cm}{!}{%
    \begin{tabular}{|c|c|c|c|c|c|c|c|c|c|c|c|c|c|c|c|c| }
\hline

T\# &     $C_0$&   $C_1$&   $C_2$&   $C_3$&   $C_4$&  $ C_5$&   $C_6$&  $ C_7$&   $C_8$&  $ C_9$&   $C_{10}$&     \small{EPOP}&   \small{OPOP}&   $R_{o}$&  POP&    $R_{p}$\\ \hline
\multicolumn{17}{|c|}{D4, D5}\\\hline

 1&       3&     25&     21&     29&     37&     37&     21&     29&     38&     33&     27   &   0.973& 0.99 &	       Y&  0.237&    Y\\\hline
 2&       1&     19&     34&     28&     29&     38&     30&     33&     28&     28&     32   &   0.973& 0.997&	       Y&  0.651&    Y\\\hline
 3&       1&     23&     30&     27&     26&     36&     31&     38&     37&     29&     22   &   0.973& 0.997&	       Y&  0.419&    Y\\\hline
 4&      4 &    30 &    28 &    30 &    21 &    35 &    38 &    23&     31 &    37 &    23   &   0.973&  0.987&	       Y&  0.258&    Y\\\hline
 5&       6&     26&     34&     28&     27&     33&     25&     28&     27&     36&     30   &   0.973&  0.98&	       Y&  0.92&    Y\\\hline
 6&       2&     28&     33&     36&     23&     25&     19&     38&     43&     30&     23   &   0.973&  0.993&	       Y&  0.004&    Y\\\hline
 7&       3&     29&     25&     30&     32&     19&     25&     27&     42&     35&     33   &   0.973&  0.990&	       Y&  0.202&    Y\\\hline
 8&       2&     31&     32&     32&     35&     28&     26&     32&     33&     28&     21   &   0.973&  0.993&	       Y&  0.804&    Y\\\hline
 9&      7 &    37 &    36 &    30&     23 &    29 &    28 &    32&     30&     26&     22   &   0.973&  0.977&	      Y&   0.195&    Y\\\hline
 10&      6&     26&     29&     23&     32&     30&     43&     24&     29&     28&     30   &   0.973&  0.98&	       Y& 0.443&    Y\\\hline
 11&      6&     43&     58&     61&     64&     56&     79&     52&     67&     54&     60   &  0.978 &  0.99&	       Y&  0.267&    Y\\\hline
 12&      3&     24&     21&     26&     33&     29&     49&     26&     32&     21&     36   &   0.973&  0.99&	       Y& 0.015&    Y\\\hline
 13&      8&     46&     61&     55&     63&     55&     59&     56&     65&     57&     75   &  0.978 &  0.987&	       Y& 0.72&    Y\\\hline
 14&     21&    223&    236&    213&    258&    245&    233&    225&    285&    246&    215  &0.984&  0.991&	       Y& 0.05&    Y\\\hline
 15&     49&    445&    545&    517&    534&    565&    534&    550&    583&    551&    527  &0.986& 0.991&	       Y& 0.320&    Y\\\hline
\multicolumn{17}{|c|}{D6, D7}\\\hline

 1&      2&    24&    25&    25&    28&    35&    38&    25&    24&    39&    35   &    0.973&  0.993& Y &   0.285 &	Y\\\hline
 2  &    1&    28&    36&    24&    31&    27&    29&    30&    39&    28&    27   &    0.973& 0.997 &Y &  0.747&	Y\\\hline
 3&      6&    23&    37&    27&    29&    37&    23&    31&    22&    38&    27   &    0.973& 0.98 & Y &  0.361&	Y\\\hline
 4&      2&    19&    27&    29&    32&    27&    29&    40&    25&    33&    37   &    0.973& 0.993 &Y &  0.384&	Y\\\hline
 5&      5&    31&    28&    32&    40&    34&    27&    27&    18&    34&    24   &    0.973& 0.983 & Y &  0.188&	Y\\\hline
 6&      2&    37&    44&    32&    17&    27&    28&    36&    30&    24&    23   &    0.973& 0.993 &Y &  0.022&	Y\\\hline
 7&     4&    30 &   28 &   29 &   28 &   26 &   24&    30 &   32&    36 &   33   &    0.973& 0.987 & Y & 0.898&	Y\\\hline
 8&     3&    35&    25 &   28 &   25&    28 &   32&    29 &   28&    25 &   42  &    0.973&  0.99 &Y &  0.35&	      Y  \\\hline
 9&     3&    28&    40 &   38 &   23&    24 &   27&    35&    26&    29 &   27   &    0.973&  0.99 & Y &  0.324&	Y\\\hline
 10&     3&    25&    27&    27&    33&    35&    29&    35&    31&    26&    29  &    0.973&  0.99 & Y &  0.95&	Y\\\hline
 11&     4&    41&    60&    48&    62&    66&    60&    65&    75&    64&    55   &    0.984& 0.993 & Y &  0.232&	Y\\\hline
 12&     1&    23&    30&    23&    29&    35&    25&    28&    45&    32&    29   &    0.973& 0.997&  Y &  0.195&	Y\\\hline
 13&     1&    69&    72&    57&    66&    56&    53&    45&    51&    68&    62   &    0.977&  0.998 &Y &  0.206&	Y\\\hline
 14&    29&   165&   215&   240&   245&   259&   258&   224&   240&   262&   263&  0.984& 0.988& Y &  0.02&	Y\\\hline
 15&    50&   444&   506&   594&   585&   560&   555&   546&   536&   507&   517 & 0.986&  0.99 &Y & 0.022&	Y\\\hline
\end{tabular}
}
\label{table:nist2} 
\end{table}



\bibliographystyle{unsrt} 
\bibliography{moje}
\end{document}